\documentclass[%
 twocolumn,
 amsmath,amssymb,
 aps,
prb,
]{revtex4-2}

\usepackage{graphicx}
\graphicspath{{figures/}}
\usepackage{dcolumn}
\usepackage{bm}
\usepackage{hyperref}
\usepackage{xspace}
\usepackage{xcolor}
\newcolumntype{C}{>{$}c<{$}}

\newcommand{\qlz}{Q_{\textrm{LZ}}\xspace}
\newcommand{\qhy}{Q_{\textrm{HY}}\xspace}
\newcommand{\qlx}{Q_{\textrm{LX}}\xspace}
\newcommand{\olz}{\Omega_{\textrm{LZ}}\xspace}
\newcommand{\olx}{\Omega_{\textrm{LX}}\xspace}
\newcommand{\ohy}{\Omega_{\textrm{HY}}\xspace}

\usepackage{xpunctuate}
\DeclareRobustCommand\etal{\xperiodafter{\emph{et al}}}

\begin{document}

\title{Transient translation symmetry breaking via quartic-order negative light-phonon coupling at the Brillouin zone boundary in KTaO${}_{3}$}

\author{Adri\'an G\'omez Pueyo}
\affiliation{
  CPHT, CNRS, Ecole Polytechnique, IP Paris, F-91128 Palaiseau, France
}%

\author{Alaska Subedi}
\affiliation{
  CPHT, CNRS, Ecole Polytechnique, IP Paris, F-91128 Palaiseau, France
}%

\date{\today}

\begin{abstract}
  KTaO${}_{3}$ presents a rich hyper-Raman spectrum originating from two-phonon
  processes at the Brillouin zone boundary, indicating the possibility
  of driving these phonon modes using intense midinfrared laser sources.
  We obtained the coupling of light to the highest-frequency longitudinal 
  optic phonon mode $\qhy$ at the $X$ $(0,0, \frac{1}{2})$ point by first 
  principles calculations of the total energy as a function of the phonon
  coordinate $\qhy$ and electric field $E$.  We find that the energy curve as a 
  function   of $\qhy$ softens for finite values of electric field, indicating 
  the presence of $\qhy^2 E^2$ nonlinearity with negative coupling coefficient. 
  We studied the feasibility of utilizing this nonlinearity to transiently 
  break the translation symmetry   of the material by making the $\qhy$ mode 
  unstable with an intense midinfrared pump pulse.  We also considered the 
  possibility that nonlinear phonon-phonon couplings can excite the 
  lowest-frequency phonon coordinates $\qlz$ and $\qlx$ at $X$ when the $\qhy$ 
  mode is externally driven. The nonlinear phonon-phonon couplings were also 
  obtained from first principles via total-energy calculations as a function of 
  the phonon coordinates, and these were used to construct the coupled classical
  equations of motion for the phonon coordinates in the presence of an external 
  pump term on $\qhy$.  We numerically solved them for a range of pump frequencies 
  and amplitudes and found three regimes where the translation symmetry is 
  broken: i) rectification of the lowest-frequency coordinates due to large amplitude 
  oscillation of the $\qhy$ coordinate about its equilibrium position, ii) rectification of only the
  $\qhy$ coordinate without displaced oscillations of the lowest-frequency coordinates, 
  and iii) rectification of all three coordinates.
  Due to the small magnitude of the coupling constant between the $\qhy$ mode 
  and the electric field, the smallest value of the pump amplitude that manages
  to transiently break the the translation symmetry of KTaO${}_{3}$ is 270 MV/cm. 
  Such a large value of electric field will likely cause a dielectric breakdown
  of the material.  However, our paper shows that light-phonon coupling with 
  negative sign can exist in real materials and motivates the search of 
  other materials with a larger magnitude of the coupling.
\end{abstract}

\maketitle


\section{Introduction}
\label{sec:introduction}

Light-induced amorphous-crystalline phase change in Ge$_2$Sb$_2$Te$_5$ 
\cite{Ovshinsky1968} and transition to a hidden state in 
$1T$-TaS$_2$ \cite{Stojchevska2014} are notable examples of
structural control of materials using light. Recently, it has been 
realized that a cubic-order nonlinear coupling between a fully-symmetric Raman
and an infrared phonon mode at the Brillouin zone center can be used to 
displace the crystal structure of a material along the
Raman phonon coordinate when the infrared phonon mode is externally 
driven \cite{Forst2011}.  
First principles calculation of nonlinear phonon couplings has been used to
propose ultrafast switching of ferroelectrics using this mechanism \cite{Subedi2015}.
Subsequent experiments have observed pump-induced transient reversal of 
electrical polarization \cite{Mankowsky2017,Henstridge2022}, although it 
has not been clarified whether this is caused by a long-period oscillation 
of a soft phonon mode or its oscillations at a displaced position.
Nonetheless, recent theoretical studies support the nonlinear phonon coupling 
as a mechanism to explain this transient reversal of ferroelectricity \cite{Mertelj2019,Abalmasov2020,Chen2022}.

Density functional theory based calculations show that quartic-order 
$Q_1^2 Q_2^2$ nonlinearity with negative coupling coefficients can 
occur between two phonon modes $Q_1$ and $Q_2$ \cite{Subedi2014}, and 
this has been used to show that a light-induced 
transition to a transient ferroelectric state in paraelectric 
materials is possible \cite{Subedi2017}. Interestingly, it has been
shown that cubic-order couplings can also be used to cause oscillations of
a symmetry-breaking Raman mode at a rectified position when it couples to a 
doubly degenerate infrared mode \cite{Radaelli2018}. There is experimental 
indication of a transient symmetry breaking in Bi$_2$Se$_3$ after 
a midinfrared pump \cite{Melnikov2020}.
Furthermore, several theoretical studies have highlighted coupling of magnetism
to phonon coordinates of a material 
\cite{Gu2016,Juraschek2017b, Fechner2018, Khalsa2018,Gu2018,Rodriguez2022,Juraschek2022,Feng2022}, 
and there are multiple experimental studies that show light-driven phononic alteration of magnetic 
behavior \cite{Nova2017, Disa2020, Stupakiewicz2021, Afanasiev2021, Disa2023}. 
The aforementioned studies have explored novel pathways for structural control of 
materials via nonlinear phononics, but they do not involve changes in the 
size of the crystalline unit cell caused by breaking the translation symmetry
of the material. 

In our previous work~\cite{Adrian2022}, we studied a method to transiently break
the translation symmetry of a crystal by pumping the highest-frequency transverse
optic (TO) phonon mode at the Brillouin zone boundary via a second-order Raman 
process.  We found that the pumped mode can soften the lowest-frequency phonon mode 
also at the zone boundary due to a quartic-order phonon-phonon coupling with 
negative coupling constant.  When the
amplitude of the oscillations of the pumped mode was large enough, the 
lowest-frequency mode became unstable, thereby breaking the translation 
symmetry.  The role of light in this symmetry-breaking process was limited to the
excitation of the highest-frequency TO phonon.

In this paper, we investigate the possibility of transiently breaking the
translation symmetry of a material by softening the pumped phonon mode itself.
Our first principles calculations show a $\alpha \qhy^2 E^2$ nonlinear 
light-phonon coupling with negative coupling constant between the 
highest-frequency longitudinal optic (LO) mode $\qhy$ of KTaO$_3$ and external 
electric field $E$, which implies that it is in principle possible to break the
translation symmetry of this material by pumping this mode.  To identify the 
peak electric field of the pump pulse required to cause the symmetry breaking, 
we numerically solved the coupled classical equations of motion of the $\qhy$ 
and lowest-frequency transverse acoustic (TA) phonon coordinates
$\qlz$ and $\qlx$.  The equations of motion were constructed using the nonlinear 
light-phonon and phonon-phonon couplings extracted from first principles 
total-energy calculations.  
The numerical solutions of these equations yield a set of dynamics of the 
phonon coordinates that is distinct from our previous study where the 
highest-frequency TO mode was pumped.
We find that the rectification of the driven LO 
coordinate $\qhy$ requires a pump amplitude of at least 370 MV/cm.  Additionally, 
the rectification of the TA coordinates
occurs when the $\qhy$ mode is pumped with a lower pump amplitude of 270 MV/cm.
These are very high values of pump intensities that are at least an order of 
magnitude larger than what can be produced using currently available midinfrared
laser sources. Furthermore, such intense sources may physically damage the samples.
Nevertheless, our paper shows that a quartic-order negative light-phonon coupling 
can occur in real materials, and this can in principle be utilized to break
the translation symmetry of a material by softening the pumped mode.

\section{Theoretical approach}
\label{sec:theory}
We are interested in the light-induced dynamics of three phonon modes of
KTaO${}_{3}$ at the $X$ $(0,\frac{1}{2},0)$ point of its Brillouin zone: 
the highest-frequency
LO mode $\qhy$ that is externally pumped and the two components of
the lowest-frequency doubly degenerate TA mode $\qlz$ and $\qlx$. The
nonlinear light-phonon and phonon-phonon couplings were obtained from 
density functional theory based first principles calculations, and the
light-induced dynamics of the system was studied using the methodology 
presented in Ref.~\cite{Subedi2014}.  This approach requires the calculation
of the phonon eigenvectors, which are then used to compute the total-energy
surface $V(\qhy,\qlx,\qlz)$ as a function of the LO and TA phonon 
coordinates (see Ref.~\cite{Subedi2021} for a review).  The phonon 
anharmonicities and phonon-phonon nonlinear couplings are then obtained
by fitting the total-energy surface with a polynomial.
We used the approach previously used by Cartella \etal~\cite{Cartella2018} 
to obtain the coupling between the pumped phonon mode and the laser pulse by
calculating the total energy as a function of the $\qhy$ mode and fitting it
with a polynomial. The phonon anharmonicities, the phonon-phonon nonlinear 
couplings, and the light-phonon coupling are used to construct the coupled 
equations of motion for the phonon coordinates in the presence of an external 
force term on the highest-frequency $\qhy$ mode. These are then solved 
numerically to obtain the structural evolution of the material as a function 
of time.

First principles calculations of the phonon frequencies and eigenvectors 
and the total-energy surfaces as a function of the phonon coordinates and 
electric field were done using the {\sc quantum espresso}~\cite{QE} (QE) 
package.  These were performed within the PBEsol generalized gradient 
approximation~\cite{PBEsol} using the ultrasoft pseudopotentials with the
valence orbitals $3s^{2}3p^{6}4s^{1}$ (K), $5s^{2}5p^{6}5d^{3}6s^{1}$
(Ta), and $2s^{2}2p^{4}$ (O) from the GBRV library~\cite{GBRV}. The plane-wave 
cutoffs for the basis set and charge density expansions were set to 60 and 
600 Ry, respectively.  We used the relaxed lattice parameter of 
$a=3.98784$ \AA\ in our calculations.
The phonon frequencies and eigenvectors at the Brillouin zone boundary point 
$X$  were calculated using density functional
perturbation theory~\cite{Savrasov1994} as implemented in QE.  The
computation of the dynamical matrix requires a previous self-consistent field
calculation, which was performed using an $8\times8\times8$ Monkhorst-Pack
$k$-point grid.

We then used the calculated phonon eigenvectors to generate modulated
structures as a function of the $\qhy$, $\qlx$, and $\qlz$ coordinates in
$1\times2\times1$ supercells and calculated their total energies
to extract the phonon anharmonicities and phonon-phonon nonlinear
couplings.
For the
total-energy surfaces calculated as a function of two phonon coordinates, we sampled
values ranging from $-2.4$ to 2.4 \AA$\sqrt{\textrm{u}}$ with a step size of 
0.08 \AA$\sqrt{\textrm{u}}$ for the TA phonon coordinates $\qlx$ and $\qlz$,
and from $-1.0$ to 1.0 \AA$\sqrt{\textrm{u}}$ with a step size of 0.05 
\AA$\sqrt{\textrm{u}}$ for the LO coordinate $\qhy$. 
In the calculations of the total-energy surface as a function of the three phonon
coordinates, we sampled values of the TA
coordinates ranging from $-3.0$ to 3.0 \AA$\sqrt{\textrm{u}}$ with a step size
of 0.1 \AA$\sqrt{\textrm{u}}$, and for the LO coordinate values from $-1.0$ to 1.0
\AA$\sqrt{\textrm{u}}$ with a step size of 0.05 \AA$\sqrt{\textrm{u}}$. These
values were chosen to modify the shortest distance between atoms of the crystal at most
10\% of their original values, allowing us to explore the anharmonicities of the material
while remaining below the Lindemann stability limit~\cite{Sokolowski2003}.
An $8\times 4\times 8$ Monkhorst-Pack $k$-point grid was used in these calculations.  
The calculated total-energy surfaces were  fit with polynomials having only the 
symmetry-allowed nonlinear terms using the 
{\sc glm}~\cite{GLM} package as implemented in {\sc julia}.  Thus obtained 
phonon anharmonicities and phonon-phonon couplings are given in 
Appendix~\ref{sec:appendix1}.

The modern theory of polarization~\cite{Souza2002} as implemented in QE
was used to calculate the total-energy surface of KTaO${}_{3}$ as a function of the
highest-frequency $\qhy$ coordinate and electric field $E$.  
We sampled the electric field using values ranging from $-14.54$ to 14.54 MV/cm 
with a step of 3.635 MV/cm and $\qhy$ ranging from $-1.0$ to 1.0
\AA$\sqrt{\textrm{u}}$ with a step of 0.1 \AA$\sqrt{\textrm{u}}$.   A slightly
denser $8\times8\times8$ Monkhorst-Pack $k$ grid was used to sample the 
Brillouin zone in these calculations.
We then fit the resulting energy surface to the following polynomial:
\begin{equation}
    \label{eq:ph-E}
    \begin{split}
    H(\qhy,E) &= \frac{1}{2}\ohy^{2}\qhy^{2} + d_{4}\qhy^{4} + d_{6}\qhy^{6} + d_{8}\qhy^{8} \\
	& \quad + rE + sE^{2} + tE^{4} + \alpha \qhy^{2}E^{2},    
    \end{split}    
\end{equation}
where the frequency $\ohy$ and anharmonic coefficients $d_{i}$ of the $\qhy$ mode 
are those extracted from the previous total-energy calculations and 
$s=-1.5386$ e\AA${}^{2}$/V, $t=-0.258$ e\AA${}^{4}$/V${}^{3}$, and $\alpha = -0.205$ e/(V u)
are the coefficients for the terms allowed by symmetry between 
the electric field and the $\qhy$ coordinate.  We get a finite value of 
$r=-259.21$ e\AA\ for the coefficient of $E$ in  $H(\qhy,E)$, which occurs due 
to the use of periodic boundary conditions.
To validate this method of obtaining light-phonon coupling, we noted in our previous 
work that the coupling of the electric field to the highest-frequency phonon of KTaO${}_3$
at the Brillouin zone center 
$\Gamma$ extracted using this method agrees well with that obtained using the 
perturbation theory approach, which is implemented in density functional theory codes 
for phonons at $\Gamma$ \cite{Adrian2022}. 
We are also aware that the largest electric field that we have used in the total-energy 
calculations is more than an order of magnitude smaller than the values that cause 
rectification of the phonon coordinates in the numerical solution of the equations of motion.  
Larger values of the electric field caused oscillations of the total energy 
during the self-consistent field iterations of density functional calculations.  
This is a limitation of the currently available computational method.  

The {\sc differentialequations}~\cite{DifferentialEquations} package from the
{\sc julia} language was used to integrate the coupled differential equations 
of motion, which was carried out using the strong stability preserving method 
of Ruuth.
The time range for the propagation was from 0 to 8 ps, and the peak amplitude 
of the laser pulse was reached at 4 ps. The initial conditions were set such that  
$\qhy=\qlx=\qlz=0.1$ \AA$\sqrt{\textrm{u}}$ and their first derivative 
with respect to time equal to zero.  We added a stochastic term in the form of 
white noise from the start of the propagation until the peak of the pump pulse 
to simulate the thermal fluctuations of the phonons~\cite{Caprini2023,Juraschek2020}. To implement
this, we picked a random value from a flat distribution going from $-1$ to 1 and then multiplied it by a
scaling factor to make it comparable to the amplitude of the initial conditions for each
mode. This scaling factor is different for the high- and low-frequency modes, as
their damping coefficients are also different.
For $\qhy$ we used 300/$N$ and for $\qlx$ and $\qlz$ 100/$N$, where $N=10^6$ is the
number of steps in the propagation.
We added this number to the corresponding phonon coordinate at the end of
the propagation step.
The presence of the stochastic term will cause the solutions of the equations of 
motion to be dependent on the random values generated for each run.
We found that solving the equations of motion five times
for the same pump amplitude and frequency was enough to asses the probability
of each type of solution, and thus we picked as the solution the one that
appeared at least three times for each pair of pump amplitude and frequency values.

In the solutions of the equations of motion, we need to distinguish the 
translation symmetry breaking 
due to rectification of the zone-boundary phonon modes from their 
long-time-period oscillations.  Furthermore, since our pump pulses have 
a finite duration, the dynamics of the phonon coordinates eventually 
revert to oscillations about their initial equilibrium positions due to the 
presence of damping terms in the equations of motion.  Our criterion 
for the breaking of translation symmetry is the presence of at least
two peaks in the oscillations of $\qhy$, $\qlx$, or $\qlz$ about a
displaced position. 

\section{Results and Discussion}
\label{sec:results}

\begin{figure}
	\centering
	\begin{minipage}{1.0\linewidth}
		\includegraphics[width=\linewidth]{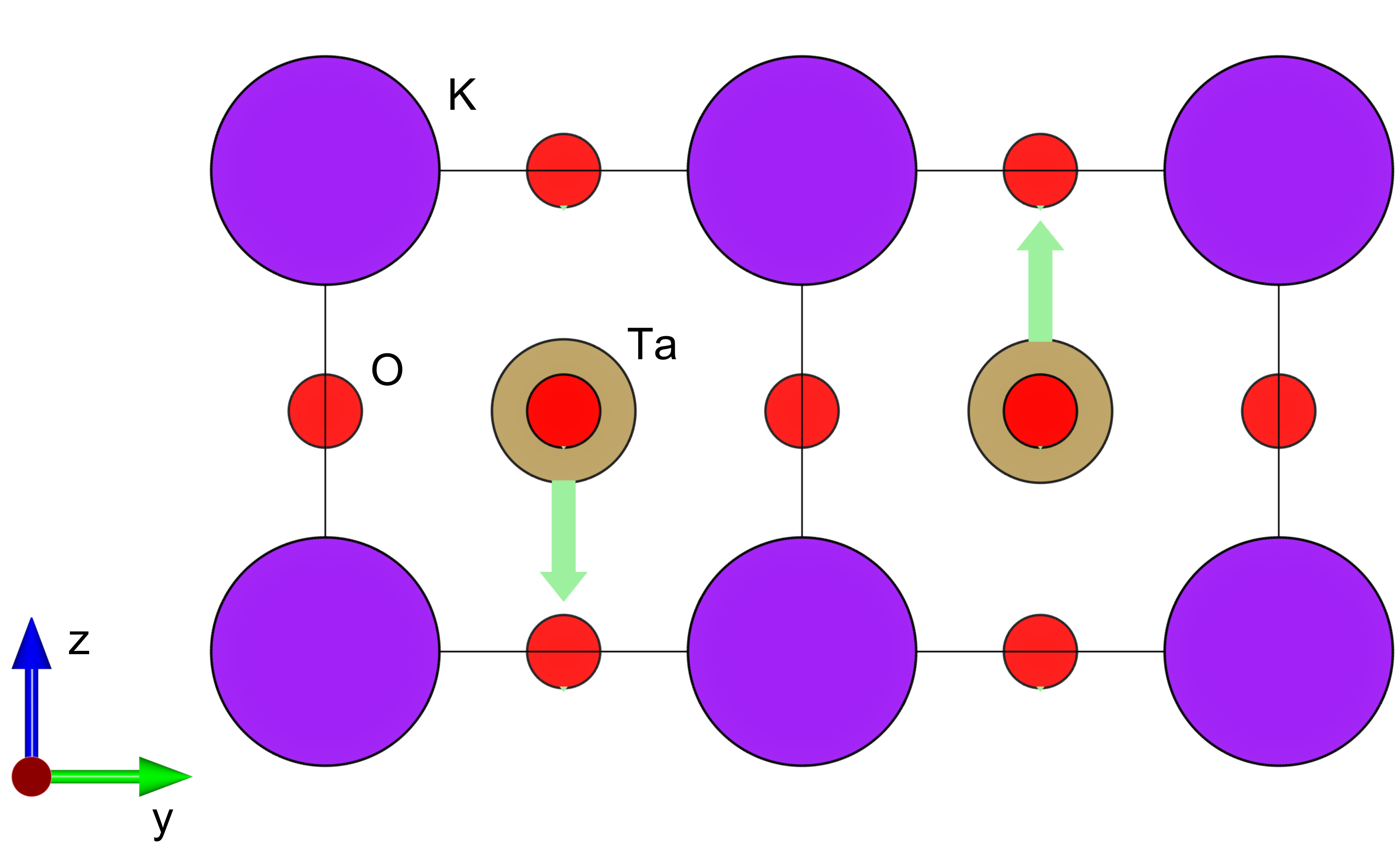}
	\end{minipage}
	\begin{minipage}{1.01\linewidth}
		\includegraphics[width=\linewidth]{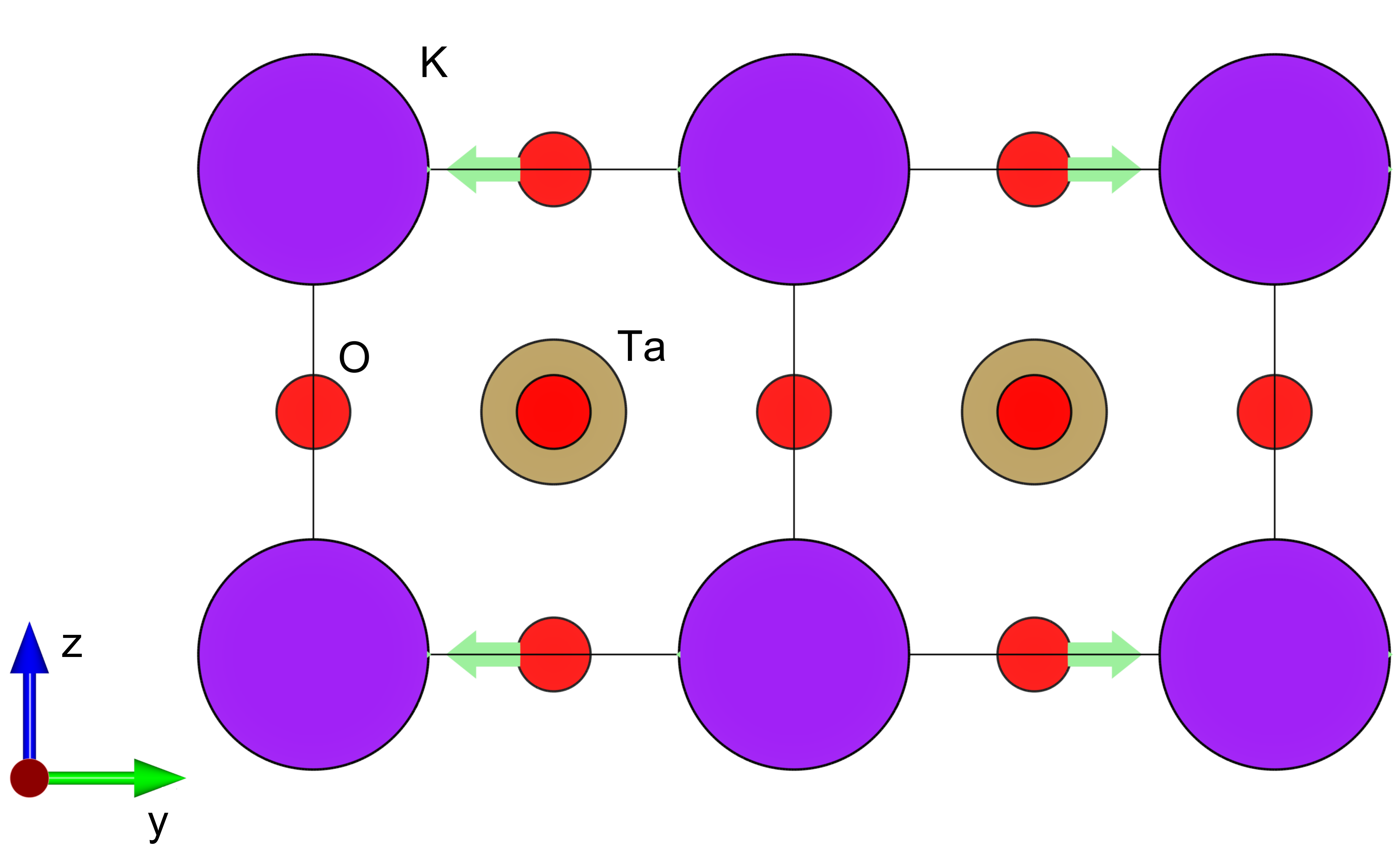}
	\end{minipage}
        \caption{Atomic displacements due to the phonon modes of KTaO$_3$ at the
        $X$ $(0,\frac{1}{2},0)$ point considered in the present paper.
        (Top) The lowest-frequency TA mode component $\qlz$ that moves Ta ions along the
        $z$ direction. The other degenerate component of this mode $\qlx$ has the 
        same atomic movements but is directed along the $x$ axis.
        (Bottom) The highest-frequency LO mode $\qhy$ that causes atomic 
        movements along the $y$ direction.}
\label{fig:phononY} 
\end{figure}

The lowest-frequency mode of KTaO$_3$ at $X$ is doubly degenerate. 
(Check Ref.~\cite{Perry1989} for the phonon band structure of KTaO$_3$.)
We follow the 
notation used in Ref.~\cite{Nilsen1967} and call it a TA mode. At a slightly 
higher frequency, there is a nondegenerate mode that is denoted as longitudinal 
acoustic. Additionally, there are four doubly degenerate TO and four nondegenerate 
LO modes, accounting for all 15 phonon modes at $X$. 
Figs.~\ref{fig:phononY}(top) and (bottom) show the atomic displacements 
corresponding to the lowest-frequency TA mode $\qlz$ and highest-frequency
LO mode $\qhy$, respectively.
The calculated frequencies of these modes are $\olx=\olz$ = 61
cm$^{-1}$ and $\ohy$ = 843 cm$^{-1}$, respectively.
They agree well with the values inferred from the Raman
experiments of Nilsen and Skinner, where these modes appear as peaks
at 123 and 1748 cm$^{-1}$ due to the doubling of the respective phonon
frequencies in second-order Raman processes \cite{Nilsen1967}. 
The two components $\qlz$ and $\qlx$ of the doubly degenerate TA mode belong 
to the irreducible representation (irrep) $X^{+}_{5}$,  while the LO mode $\qhy$ 
has the irrep $X^{-}_{3}$.  
The $\qlz$ mode involves displacement of the Ta ions
against the O octahedra along the $z$ direction. The $\qhy$ mode displaces
the pair of O ions situated on the faces of the unit cell parallel to
the $xy$ plane against the K ion in the $y$ direction (though
the displacement of the K ion is so small that it is not perceptible in the
figure), while the rest of the O ions remain stationary.
The atomic displacements within the adjacent unit cells are out of phase 
along the $y$ direction because these modes have the wave vector $(0,\frac{1}{2},0)$.
They thus break the translation symmetry of the crystal. 

\begin{figure}[htp]
	\includegraphics[scale=0.7]{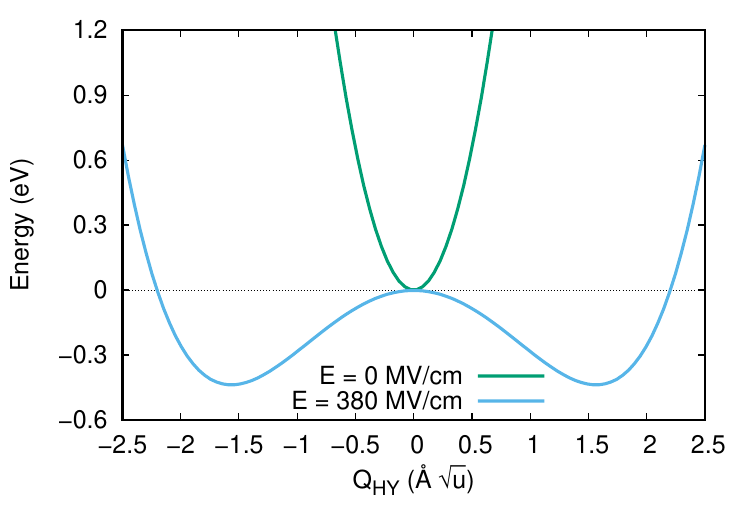}
	\caption{\label{fig:phonon-light} Total energy as a 
        function of the highest-frequency $\qhy$ phonon coordinate for
        electric field $E = 0$ and $380$ MV/cm.}
\end{figure}

Previously, we found that the energy curve of the highest-frequency TO mode at 
$X$ of KTaO$_3$ stiffens in the presence of a finite electric field, indicating
the presence of a $\qhy^2 E^2$ nonlinear coupling with a positive coefficient 
between the phonon coordinate
and electric field \cite{Adrian2022}. When we performed similar total-energy 
calculations as a function of the highest-frequency LO mode $\qhy$ for different 
values of the electric field $E$, we found a softening of the energy curve
of the phonon coordinate.  This softening is symmetric with respect to the
sign of both the phonon coordinate ($\qhy\rightarrow-\qhy$) and the electric 
field ($E\rightarrow-E$), which shows the presence of a $\qhy^2 E^2$ nonlinear
coupling with a negative coupling coefficient $\alpha$.  
The presence of a $\qhy^{2}E^{2}$ nonlinearity is consistent with the 
fact that this is the lowest-order coupling term between an electric field and 
finite-wave-vector phonon allowed by symmetry~\cite{Bartels2000}.  However, 
the negative sign of the coupling coefficient $\alpha$ is surprising.
A fit of the total-energy surface $H(\qhy,E)$ by a polynomial gives a relatively 
small value 
for the coupling constant $\alpha=-0.205$ e/(V u).  Although the magnitude of the 
coupling constant is small, this in principle makes it possible to break the 
translation symmetry of the crystal by destabilizing the pumped phonon mode itself.  
This is illustrated by Fig.~\ref{fig:phonon-light}, which shows the development of
symmetrical minima at $\qhy\neq0$ for an extrapolated electric field of 
$E = 380$ MV/cm.  In order to break the translation symmetry for
values of the electric field around 10 MV/cm, we need $\alpha\approx150$ e/(V u).
When the $\qhy$ coordinate is externally pumped, it experiences
a force $-\partial H/\partial\qhy$ that leads to a pump-induced renormalization 
of its frequency $\ohy^2 \rightarrow \ohy^2 (1 + 2 \alpha E^2)$. Since the correction
due to the electric field has an even power, the softening is not averaged out 
over time, and the $\qhy$ coordinate becomes unstable for $E > \frac{1}{\sqrt{2\alpha}}$.

\begin{figure}[htp]
	\includegraphics[scale=0.7]{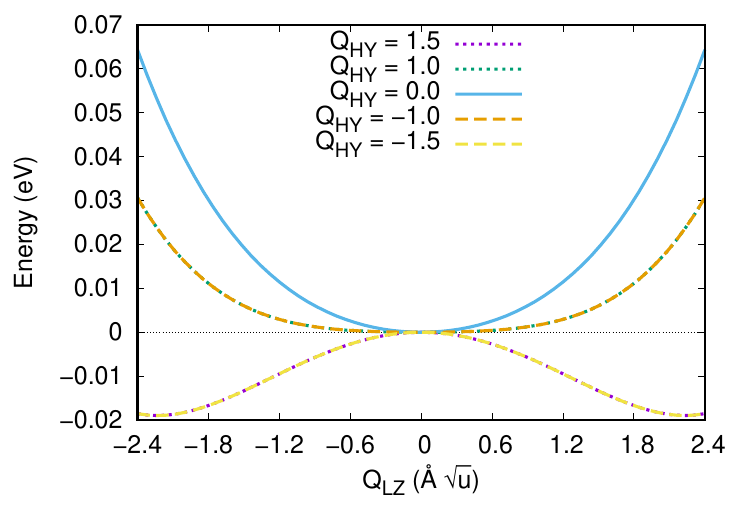}
	\caption{\label{fig:double_well} Calculated total energy as a function
          of the lowest-frequency TA coordinate $\qlz$ for different values of 
          the highest-frequency LO coordinate $\qhy$. For visual clarity, 
          the zero energy point has been chosen so that the curves coincide 
          at $\qlz=0$.}
\end{figure}

The pumped energy to the $\qhy$ coordinate should also flow to the 
lowest-frequency phonon mode at $X$. To make our paper more realistic,
we include the dynamics of the $\qlz$ and $\qlx$ components of the 
TA mode in our 
simulations.
Fig.~\ref{fig:double_well} shows
five energy curves from the calculated energy surface $V(\qhy,\qlx=0,\qlz)$.  
We can again see that the energy curves are symmetric upon the 
transformations $\qlz \rightarrow -\qlz$ and $\qhy \rightarrow -\qhy$, 
which implies that the energy surface is an even function of both $\qlz$ and
$\qhy$. Therefore, only terms with even powers of these coordinates occur
in the polynomial fit of the total-energy surface.  The fact that these
modes belong to different irreps imposes this constraint. 
Since both the $\qlx$ and $\qlz$ TA components have the same
irrep, the same reasoning can be applied to the energy surface 
$V(\qhy,\qlx,\qlz=0)$ and its polynomial fit.

The energy curve of the $\qlz$ coordinate softens and develops 
a double-well shape as the magnitude of the $\qhy$ coordinate
is increased.  Accordingly, the fit of $V(\qhy,\qlx=0,\qlz)$
yields negative sign for the coefficients of the nonlinear 
coupling terms $h_1 \qhy^2 \qlz^2$, $h_2 \qhy^4 \qlz^2$, and
$h_4 \qhy^6 \qlz^2$ (see Appendix \ref{sec:appendix1}). 
The total force experienced along the $\qlz$ coordinate is given 
by $-\partial V/\partial\qlz$, and the nonlinear terms cause a 
renormalization of its frequency as $\olz^2 \rightarrow \olz^2 (1 + 2 h_1
\qhy^2 + 2 h_2 \qhy^4 + 2 h_4 \qhy^6 + \cdots)$.  Only even powers
of $\qhy$ appear in this correction.  Hence, the softening of the frequency of 
the $\qlz$ coordinate due to the oscillation of the $\qhy$ coordinate 
will not be averaged out over time. The same softening should also occur for 
the $\qlx$ coordinate because both components of the lowest-frequency TA mode 
have the same irrep.

We used the calculated total-energy surfaces as the potential energy of the
$\qhy$, $\qlx$, and $\qlz$ coordinates and used them to construct their 
coupled equations of motion, which read 
\begin{align}
  \label{eq:eq_motion}
  \ddot{Q}_{\textrm{HY}} + \gamma_{\textrm{HY}}\dot{Q}_{\textrm{HY}} + \ohy^2 \qhy &= -\frac{\partial V^{{\textrm{nh}}}(\qhy, \qlx, \qlz)}{\partial \qhy} \nonumber\\
   & \quad + F(t), \nonumber \\
  \ddot{Q}_{\textrm{LX}} + \gamma_{\textrm{LX}}\dot{Q}_{\textrm{LX}} + \olx^2 \qlx &= -\frac{\partial V^{{\textrm{nh}}}(\qhy, \qlx, \qlx)}{\partial \qlx},\nonumber\\
  \ddot{Q}_{\textrm{LZ}} + \gamma_{\textrm{LZ}}\dot{Q}_{\textrm{LZ}} + \olz^2 \qlz &= -\frac{\partial V^{{\textrm{nh}}}(\qhy, \qlx,
   \qlz)}{\partial \qlz}.    
\end{align}
Here $V^{{\textrm{nh}}}(\qhy,\qlx,\qlz)$ is the nonharmonic part of
the polynomial fit to the calculated total-energy surfaces as a
function of the three coordinates. Its full expression is given in 
Appendix~\ref{sec:appendix1}. The damping coefficients $\gamma_{i}$
are set to 10\% of the value of their corresponding natural frequency.
The external force $F(t)$ experienced by the $\qhy$ coordinate due to
its coupling with the electric field of the pump pulse is given by
\begin{equation}
\label{eq:lph-coupling}
        F  =  - \frac{\partial H(\qhy,E)}{\partial \qhy} 
           =  - 2 \alpha \qhy E^2. 
\end{equation}
Since pulsed laser sources with finite time duration are used in most 
pump-probe experiments, 
we studied the dynamics using Gaussian-enveloped single-frequency
pulses for the electric field
\begin{align}
\label{eq:pumpsf}
E_{\rm sf}(t) & = E_{0}\sin(\omega t)e^{-t^{2}/2(\sigma/2\sqrt{2\log
    2})^{2}}.
\end{align}
Here $E_0$, $\omega$, and $\sigma$ are the amplitude, frequency, and 
duration (full width at half maximum) of the pulse, respectively.  
We used $\sigma = 2$ ps in all our simulations. Chirped laser sources are 
also used in pump-probe experiments, and we have
repeated our simulations with Gaussian-enveloped chirped pulses. These
results are presented in Appendix~\ref{sec:chirped-pulse}, as they were 
analogous to those obtained with the single-frequency pulse but with a 
slightly lower threshold for the pulse amplitude that manages to achieve
rectification of the phonon coordinates.

For a given value of pump frequency, we solved the coupled equations of 
motion of the $\qhy$, $\qlx$, and $\qlz$ coordinates given in 
Eq.~\ref{eq:eq_motion} for a range of pump amplitude $E_0$.  When the pump 
amplitude is small, the pumped mode $\qhy$ oscillates at its natural frequency 
$\ohy$ without getting amplified and decays at a rate determined by 
$\gamma_{\textrm{HY}}$. Hence, the energy
transferred to the $\qlz$ and $\qlx$ coordinates is also small, and they
also briefly oscillate at their natural frequencies $\olz = \olx$ without 
getting amplified. At the other extreme, all three modes diverge at very 
large values of the the pump amplitude, which corresponds to the 
dielectric breakdown of the material. In between these uninteresting scenarios, 
we searched for a range of pump frequency and amplitude where
at least one of the three coordinates oscillates at a displaced position.

\begin{figure}[htp]
	\includegraphics[width=\columnwidth]{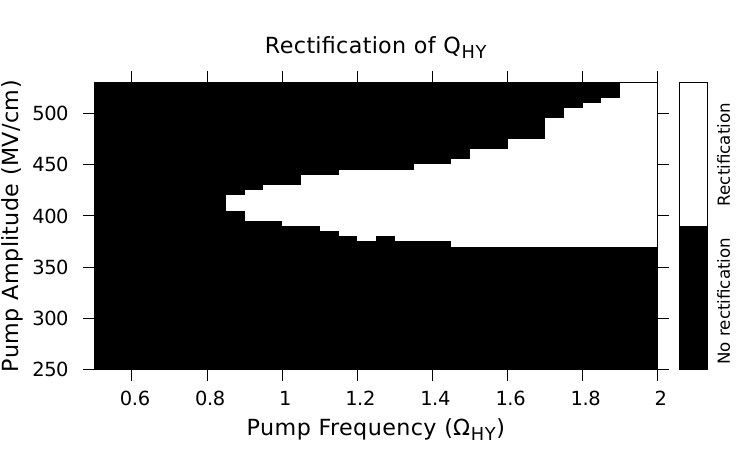}
	\includegraphics[width=\columnwidth]{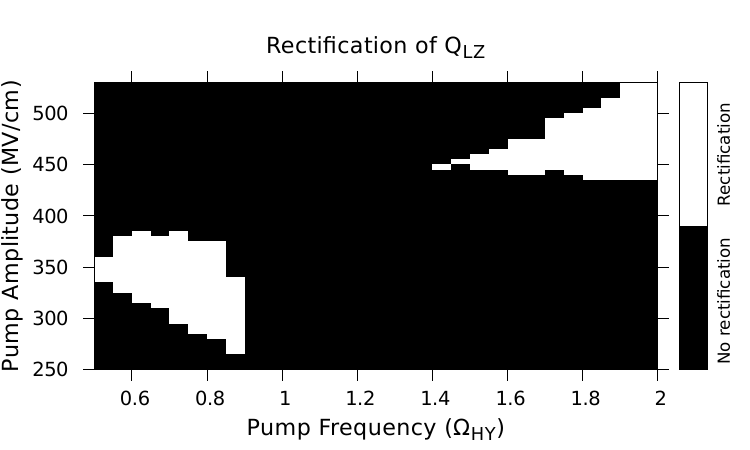}
	\caption{\label{fig:rect-sm} Pump amplitudes and frequencies
	of the single-frequency pulse driving the $\qhy$ phonon coordinate
	that induce rectification of the $\qhy$ (top) and $\qlz$ (bottom)
	coordinates. As both components of the lowest-frequency TA mode have the same irrep,
    the results for the $\qlx$ coordinate are analogous to those of $\qlz$.}
\end{figure}

\begin{figure*}[htp]
	\includegraphics[width=\textwidth]{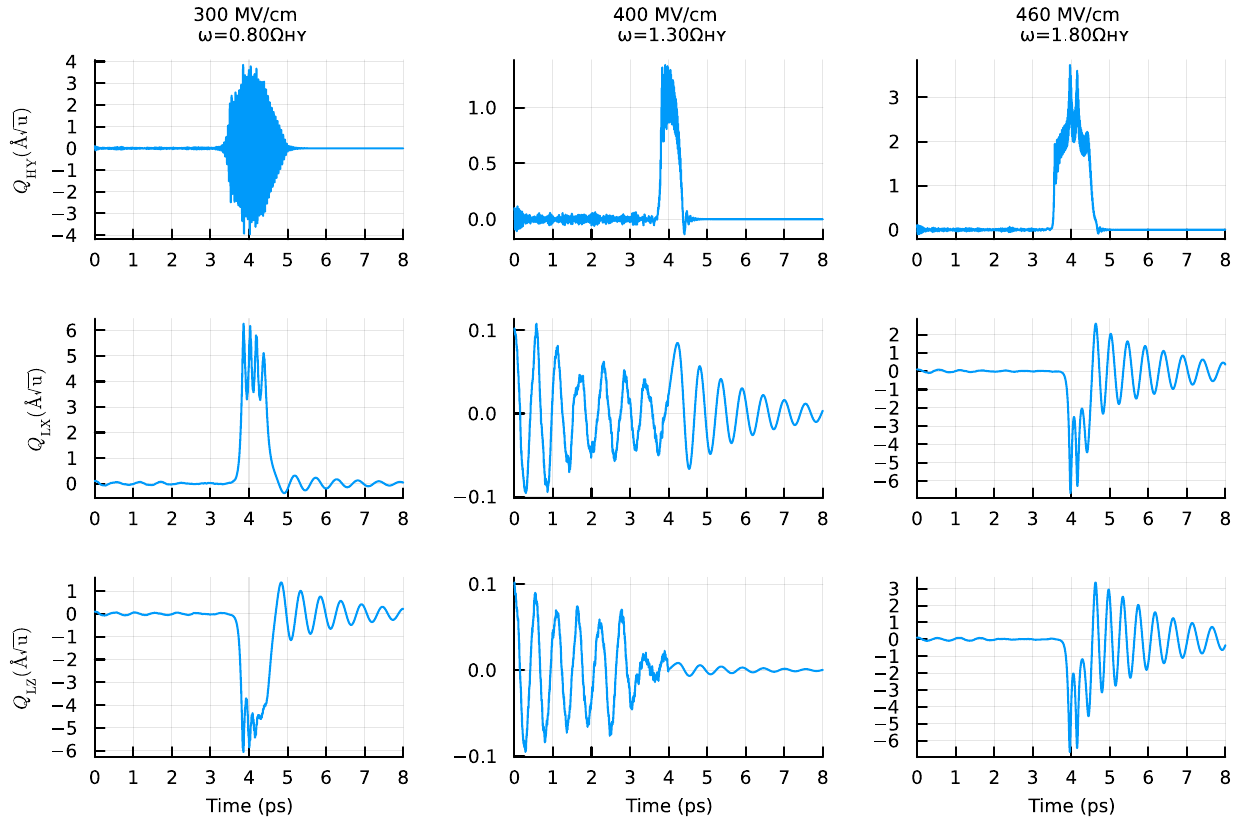}
	\caption{\label{fig:comparison-sm} Examples of the three rectification regimes 
    of the $\qhy$, $\qlx$, and $\qlz$ phonon coordinates for pump pulses with frequencies of
	$\omega=0.80\ohy$ (left), $\omega=1.30\ohy$ (center), and $\omega=1.80\ohy$
	(right) and amplitudes of 300, 400, and 460 MV/cm respectively.}
\end{figure*}

We find that for very low pump frequencies the effect of the  pump pulse is no 
different from applying a constant electric field, driving the pumped $\qhy$ 
mode close to its natural frequency $\ohy$ for the duration of the pulse. 
Above pump frequency of $\omega > 0.50 \ohy$, we were able to find pump amplitudes
for three different types of rectification of the phonon coordinates, which is 
shown in Fig.~\ref{fig:rect-sm}. For $0.50 \ohy < \omega < 0.85 \ohy$, the 
lowest-frequency TA coordinates $\qlz$ and $\qlx$ get rectified while the pumped 
$\qhy$ mode oscillates at its equilibrium position. Interestingly, in the range 
$ 0.85 \ohy < \omega < 1.40 \ohy$, the pumped mode $\qhy$ itself oscillates at a 
displaced position while the oscillations of the lowest-frequency TA coordinates occur about
their equilibrium positions. Finally, at $\omega > 1.40 \ohy$ all three coordinates get 
rectified.

The rectification of the lowest-frequency $\qlz$ and $\qlx$ coordinates for values of 
pump frequencies in the window $0.50\ohy < \omega < 0.85\ohy$ is similar to the 
behavior obtained  by us in our previous work where the highest-frequency TO mode was 
externally pumped~\cite{Adrian2022}.  The rectification occurs because they 
experience a time-averaged double-well potential while the 
high-frequency LO coordinate $\qhy$ oscillates around its equilibrium position with
a large enough amplitude.  
At a pump frequency of $\omega = 0.50 \ohy$, we find that the earliest onset of the 
rectification of the TA modes occurs for a pump amplitude of 335 MV/cm. Smaller amplitudes
do not excite the LO mode enough to induce the rectification of the TA modes, and amplitudes
larger than 350 MV/cm cause the TA modes to quickly oscillate between the two wells of the
potential with a time average of zero.
As the pump frequency is increased within the
window of this regime, the TA modes start getting rectified at lower pump amplitudes, 
while the solution of the equations of motion starts to diverge as the pump amplitude
is increased above 380 MV/cm.  As a result, the range of amplitudes that induce the 
rectification of the TA coordinates broadens with the frequency of the pump pulse until
a new regime is reached above pump frequency of $\omega > 0.85 \ohy$.  This behavior 
is reflected in the left areas  of Figs.~\ref{fig:rect-sm} (top) and (bottom).  
The left column of Fig.~\ref{fig:comparison-sm} shows the time evolutions of the
three coordinates for a pump frequency of  $\omega=0.80\ohy$ and amplitude of 300 MV/cm,
which is a representative solution of the equations of motion in this regime.  They exhibit
rectification of the TA coordinates with $\langle \qlz \rangle$ and $\langle \qlx \rangle
\neq 0$, while the LO coordinate oscillates about the equilibrium position with 
$\langle \qhy \rangle = 0$.  Surprisingly, we find that that the time-averaged position about which the TA 
coordinates oscillate at a displaced position does not change as a function of the pump 
amplitude in this regime.

\begin{figure}[htp]
	\includegraphics[width=0.5\textwidth]{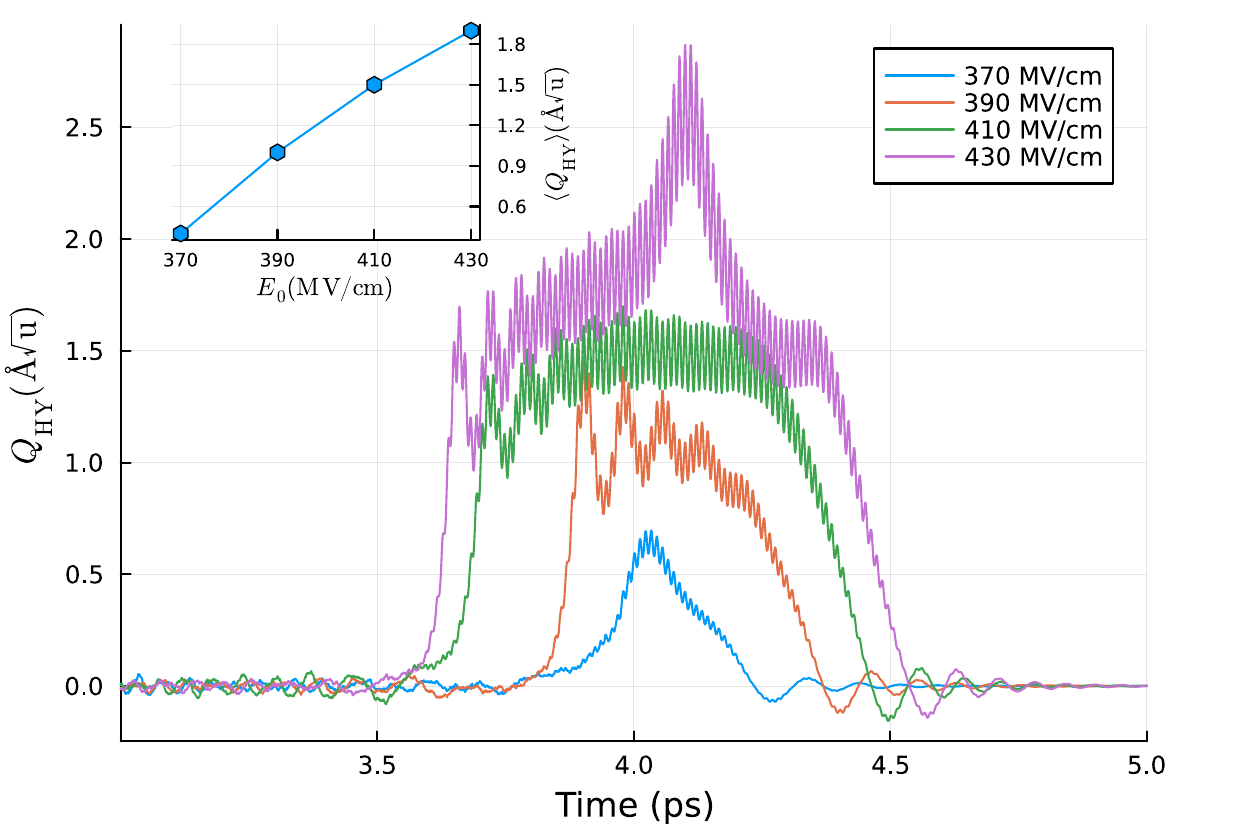}
	\caption{\label{fig:ampl-growth} Growth of the
          rectified position of the $\qhy$ phonon coordinate as a function
          of the amplitude of the pump. The frequency of the pulse
          was set to $\omega=1.80\ohy$, where $\ohy$ = 843 cm$^{-1}$. Inset: The
          average position $\langle\qhy\rangle$ during the rectification as a function
          of the peak amplitude of the applied electric field $E_{0}$. The line connecting
          the dots is a guide for the eye.}
\end{figure}

For values of pump frequency in the window $0.85 \ohy < \omega < 1.40 \ohy$,
we find a regime where the pumped LO mode $\qhy$ is rectified but the lowest-frequency
TA components remain oscillating around their equilibrium position.  
Interestingly, the amplitude of the pumped coordinate $\qhy$ drops significantly
upon entering this regime. An example of 
the time evolutions of the three phonon coordinates in this regime is shown 
in the middle column of Fig.~\ref{fig:comparison-sm}, which was obtained for a
pump frequency $\omega = 1.30 \ohy$ and amplitude $E_0 = 400$ MV/cm.  
Instead of oscillating about the equilibrium position, the pumped mode 
gets rectified because of the negative sign of the coefficient $\alpha$ of 
the quartic-order $\qhy^2 E^2$ light-phonon coupling term.
For a pump frequency $\omega = 0.85 \ohy$ that is near the beginning of this 
window, the rectification of the LO mode happens for pump amplitudes from 405 to
415 MV/cm.  
As the pump frequency is increased, the range of the pump amplitude that rectifies 
the $\qhy$ mode broadens. For a pump frequency of $\omega=1.40\ohy$,
this range goes from 375 to 440 MV/cm, as we can see on the top panel of
Fig.~\ref{fig:rect-sm}. Larger values of pump amplitude cause the equations of motion
to diverge, signaling the breakdown of the material.  However, until the 
divergence occurs, the new position
about which the rectified oscillations of the $\qhy$ mode occurs increases with the
amplitude of the pump in approximately linear fashion.
This is illustrated in Fig.~\ref{fig:ampl-growth},
which shows the time evolution of $\qhy$ for four increasing values of pump 
amplitude. This change in the displaced position of the $\qhy$ coordinate is different from the almost
unchanged rectified position of the TA coordinates discussed in the previous paragraph. 

When pump frequency is in the range $0.85 \ohy < \omega < 1.40 \ohy$,  the values reached 
by the rectified $\qhy$ 
coordinate are not large enough to induce a rectification of the TA modes 
through their phonon-phonon coupling before the divergence of the equations of motion.  
However, for pump frequencies $\omega > 1.40 \ohy$, the values reached by the pumped
coordinate $\qhy$ start to be comparable to that reached by it in the first regime, and
we enter a distinct regime where we can find pump amplitudes that can rectify the three
modes at the same time.  
At pump frequency  $\omega = 1.45 \ohy$, pump amplitudes between 370 and 440 MV/cm 
cause rectification of the
LO mode without rectifying the TA mode, as discussed above.  But when the pump amplitude is
increased above 440 MV/cm, the oscillation about the transient displaced position of the 
$\qhy$ coordinate is large enough to induce rectification of the $\qlz$ and $\qlx$ TA 
coordinates.  
This behavior has an activation threshold for the pump amplitude of 440 MV/cm,
but the maximum amplitude that manages to induce this triple rectification
without causing the breakdown of the material grows linearly with the frequency until
reaching a plateau of around 530 MV/cm starting at $\omega=1.90\ohy$. An example of 
the time evolution of the three coordinates in this regime is shown in the right
column of Fig.~\ref{fig:comparison-sm}.  One notable feature of this regime is 
that the $\qhy$ mode beats at the same frequency as the $\qlz$ and $\qlx$ modes, 
suggesting that the energy that flows to the TA coordinates dominates the dynamics of
the system.

\section{Summary and Conclusions}
\label{sec:conclusion}

In summary, we have discovered using first principles total-energy
calculations that a quartic-order $\alpha \qhy^2 E^2$ coupling with a negative 
coefficient $\alpha$ occurs between the highest-frequency LO phonon coordinate
$\qhy$ of KTaO$_3$ at the Brillouin zone boundary point $X$ and electric 
field $E$. This implies that the $\qhy$ mode softens when it is driven 
by an external laser source. We investigated the feasibility of
transiently breaking the translation symmetry of KTaO$_3$ by driving 
the $\qhy$ mode to instability using pump pulses with high electric field.  
We also 
considered the coupling of the $\qhy$ mode with  the lowest-frequency 
TA phonon modes $\qlz$ and $\qlx$ at $X$. The nonlinear couplings 
between these modes were also obtained from first principles by 
calculating the total energy as a function of the phonon coordinates. 
We find that the energy curves of the TA coordinates $\qlz$ and $\qlx$ 
develop a double-well shape for finite values of the $\qhy$ coordinate,
suggesting that these modes could also become unstable when the $\qhy$
is externally pumped. We used the calculated nonlinear couplings to 
construct the coupled equations of motion for the three coordinates in 
the presence of a Gaussian-enveloped pump pulse term on the $\qhy$ mode.  
These were then numerically solved for a range of pump frequencies and 
amplitudes.

We find three different regimes of light-induced translation symmetry 
breaking, which occur for pump frequency $\omega > 0.5 \ohy$ when a
Gaussian-enveloped single-frequency pump pulse is used. Only the TA
coordinates $\qlz$ and $\qlx$ rectify when pump frequency is in the 
range $0.5 < \omega < 0.85 \ohy$. The lowest pump amplitude that can 
cause rectification in this regime is 270 MV/cm. When pump frequency
is in the range $0.85 < \omega < 1.40\ohy$, only the pumped mode 
$\qhy$ rectifies. A pump amplitude of at least 375 MV/cm is required
to exhibit this behavior. Finally, all three coordinates can rectify
when the pump frequency is greater than $1.40 \ohy$, which
requires a pump amplitude of at least 440 MV/cm. We find that the 
use of chirped pulses only modestly decreases the threshold required
to rectify these modes.

The pump amplitude required to break the translation symmetry of 
KTaO$_3$ by pumping its highest-frequency LO coordinate is well beyond 
what can be produced by currently available experimental setups. Even
if powerful laser sources were available, such intense pump pulses would
likely cause dielectric breakdown of the sample. Nevertheless, we 
have shown that quartic-order light-phonon coupling with negative sign 
can occur in real materials that can be utilized to break the translation 
symmetry of the crystal by rectifying the pumped mode. Our paper motivates 
the search for materials where the magnitude of this light-phonon coupling 
is large.

\begin{acknowledgments}
This work was supported by the Agence Nationale de la Recherche
under grant ANR-19-CE30-0004 ELECTROPHONE and GENCI-TGCC under grant
A0110913028.
\end{acknowledgments}


\appendix

\section{\uppercase{Chirped Pulse}}
\label{sec:chirped-pulse}

\begin{figure}[htp]
	\includegraphics[width=\columnwidth]{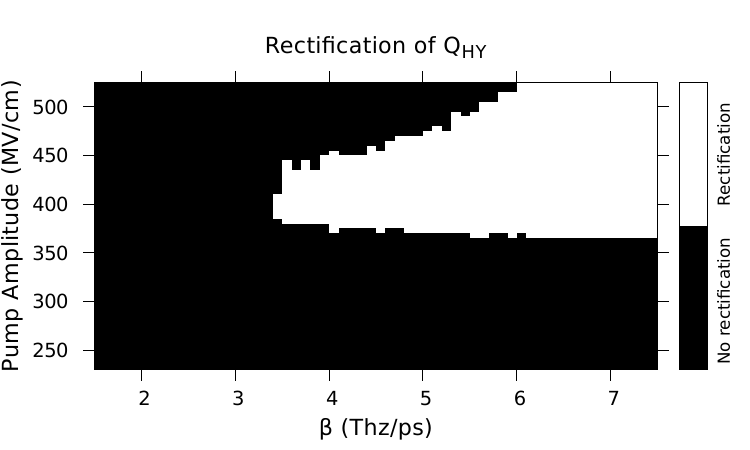}
	\includegraphics[width=\columnwidth]{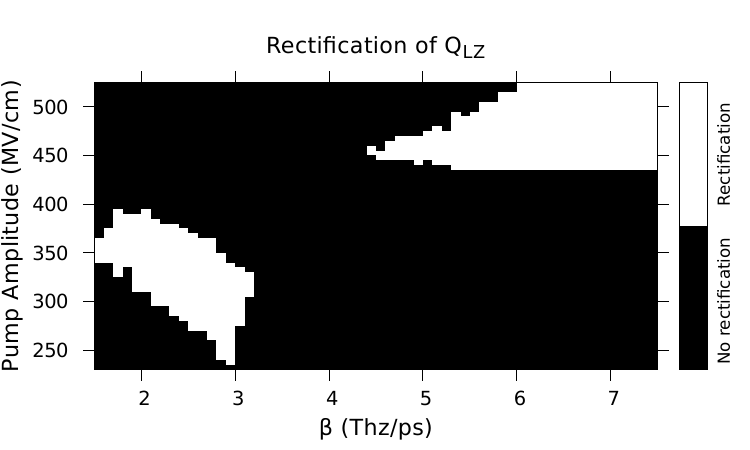}
	\caption{\label{fig:rect-chirp} Pump amplitudes and 
    frequency-growth rate parameters $\beta$
	of the chirped pulse driving the $\qhy$ phonon coordinate 
	that induce a rectification of the $\qhy$ (top) and $\qlz$ (bottom)
	coordinates. As both components of the lowest-frequency TA mode have the same irrep,
    the results for the $\qlx$ mode are analogous to those of $\qlz$.}
\end{figure}

\begin{figure*}[htp]
	\includegraphics[width=\textwidth]{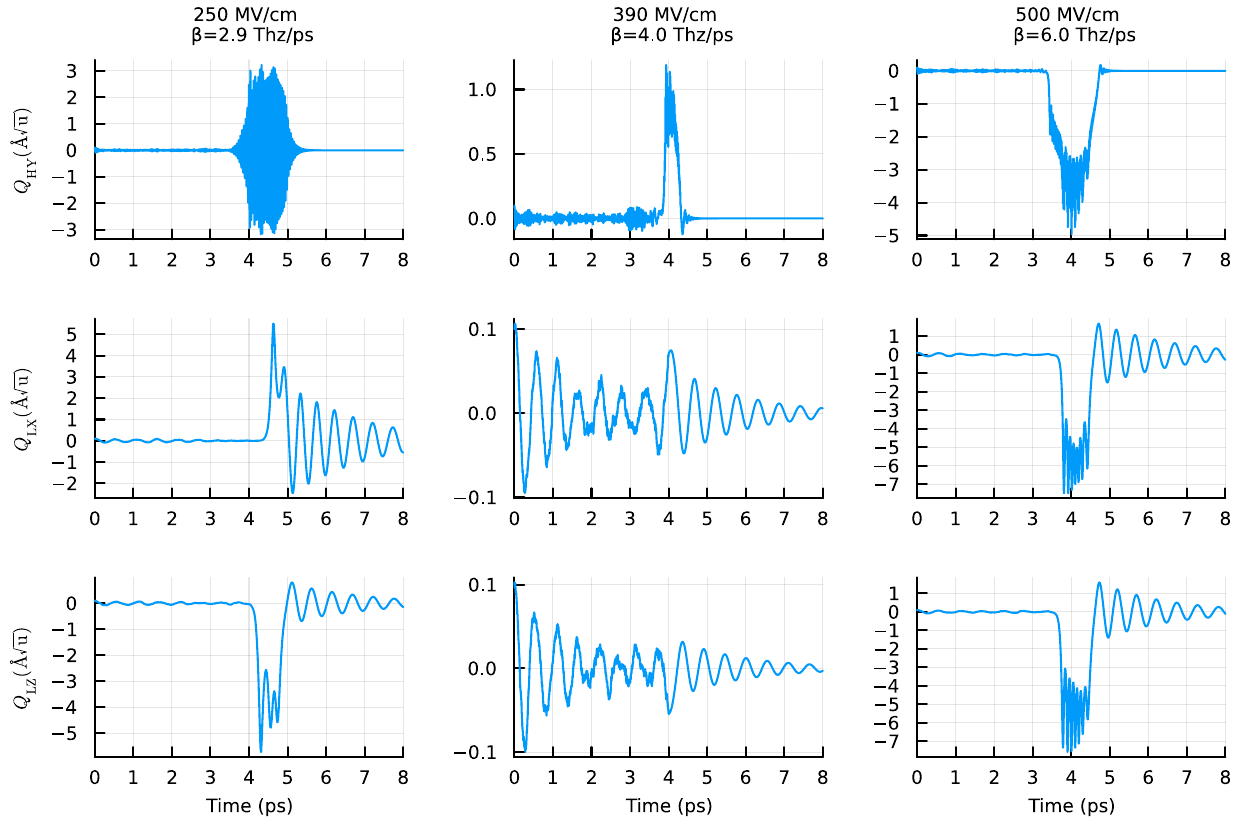}
	\caption{\label{fig:comparison-chirp} Examples of the three
    rectification regimes of the $\qhy$, $\qlx$, and $\qlz$ phonon
    coordinates for chirped pulses with frequency-growth rate parameters of  
    $\beta=2.9$ Thz/ps (left),
    $\beta=4.0$ Thz/ps (center), and $\beta=6.0$ Thz/ps
	(right) and amplitudes of 250, 390, and 500 MV/cm respectively.}
\end{figure*}

Chirped pulses can be more effective at driving the phonon modes of a material,
especially if the pumped mode transiently hardens \cite{Itin2018}.  
We also studied the light-induced dynamics of the three phonon coordinates 
described by Eqs.~\ref{eq:eq_motion} using Gaussian-enveloped chirped pulses
having electric field
\begin{align}
\label{eq:pumpch}
E_{\rm ch}(t) & = E_{0}\sin(\beta t^{2})e^{-t^{2}/2(\sigma/2\sqrt{2\log 2})^{2}}.
\end{align}
Here, $\beta$ is the chirp parameter that describes the linear growth of the 
pulse as a function of time.

Fig.~\ref{fig:rect-chirp} shows the values of the amplitude  and $\beta$ of 
the pump pulse that manage to induce the rectification of the phonon modes.
We find that the effect of the chirped pulse is similar to 
applying a constant electric field for the growth rate of the frequency $\beta<1.5$
THz/ps, the same situation we found when using single-frequency pump pulses with 
frequencies below $0.50\ohy$.
Above $\beta > 1.5$ THz/ps, we can distinguish the same three regimes that we found for the 
single-frequency pulse: (i) rectification of the lowest-frequency TA coordinates $\qlz$ and
$\qlx$,  (ii) rectification of the pumped LO coordinate $\qhy$, and (iii)
rectification of all three coordinates.  The rectification of only the TA coordinates
happens for $1.5 < \beta < 3.1$ THz/ps.  The lowest value of the pump amplitude that
causes rectification in this regime is 235 MV/cm, which occurs for $\beta = 2.9$ THz/ps.
This is slightly smaller than the value of 270 MV/cm obtained for a single-frequency 
pulse.     
For values of the chirp parameter $\beta > 3.4$ THz/ps, we can find pump amplitudes 
that rectify the pumped phonon coordinate $\qhy$.  The lowest pump amplitude that
rectifies $\qhy$ is 385 MV/cm at $\beta = 3.4$ THz/ps. This threshold decreases as 
$\beta$ is increased until $\beta = 6.1$ THz/ps, where it stabilizes at 365 MV/cm.  
For comparison, the lowest pump value to cause rectification of $\qhy$  using
single-frequency pulse is 370 MV/cm.  Only the pumped LO coordinate $\qhy$ gets 
rectified for $3.4 < \beta < 4.4$ THz/ps, whereas pump amplitudes that cause 
rectification of all three coordinates can be found for $\beta > 4.4$ THz/ps. An example of 
each regime is plotted in Fig.~\ref{fig:comparison-chirp}.

The results obtained using the chirped pump pulse are analogous to those found for 
the single-frequency pulse, except for the slightly smaller threshold value of the pump pulse 
amplitude that causes rectification of the phonon coordinates.
The efficiency of the pump increases because the chirped pulse includes a wider range of 
frequencies. The  
frequency of the pumped LO mode stiffens while it is driven to large-amplitude
oscillations due to the presence of anharmonic terms. Hence, a pulse that is capable of
matching this changing frequency is more efficient at exciting the phonon mode
than the single-frequency one.


\onecolumngrid
\section{\uppercase{Expression for the Total Energy Surface}}
\label{sec:appendix1}

As in our previous work \cite{Adrian2022}, the phonon anharmonicities and 
phonon-phonon couplings between the three coordinates $\qhy$, $\qlz$, and $\qlx$ 
were obtained by fitting the calculated total-energy surface $V(\qhy, \qlx, \qlz)$
with the following expression
\begin{equation}
    \label{eq:V}
    \begin{split}
        V & = \frac{1}{2}\olx^{2}\qlx^{2} + \frac{1}{2}\olz^{2}\qlz^{2} + \frac{1}{2}\ohy^{2}\qhy^{2}
           + V^{{\textrm{nh}}},
    \end{split}
\end{equation}
where the nonharmonic part $ V^{{\textrm{nh}}}(\qhy, \qlx, \qlz)$ is given by
\begin{equation}
\label{eq:Vnh}
\begin{split} 
        V^{{\textrm{nh}}} & = a_{4}\qlx^{4} + a_{6}\qlx^{6} + a_{8}\qlx^{8} 
        + b_{4}\qlz^{4} + b_{6}\qlz^{6} + b_{8}\qlz^{8}
        + d_{4}\qhy^{4} + d_{6}\qhy^{6} + d_{8}\qhy^{8} \\
        & \quad + e_{1}\qlx^{2}\qlz^{2} + e_{2}\qlx^{4}\qlz^{2} + e_{3}\qlx^{2}\qlz^{4}
        + e_{4}\qlx^{6}\qlz^{2} + e_{5}\qlx^{4}\qlz^{4} + e_{6}\qlx^{2}\qlz^{6}\\
        & \quad + h_{1}\qhy^{2}\qlx^{2} + h_{2}\qhy^{4}\qlx^{2} + h_{3}\qhy^{2}\qlx^{4}
        + h_{4}\qhy^{6}\qlx^{2} + h_{5}\qhy^{4}\qlx^{4} + h_{6}\qhy^{2}\qlx^{6}\\
        & \quad + i_{1}\qhy^{2}\qlz^{2} + i_{2}\qhy^{4}\qlz^{2} + i_{3}\qhy^{2}\qlz^{4}
        + i_{4}\qhy^{6}\qlz^{2} + i_{5}\qhy^{4}\qlz^{4} + i_{6}\qhy^{2}\qlz^{6}\\
        & \quad + k_{1}\qhy^{2}\qlx^{2}\qlz^{2} + k_{2}\qhy^{4}\qlx^{2}\qlz^{2}
        + k_{3}\qhy^{2}\qlx^{4}\qlz^{2} + k_{4}\qhy^{2}\qlx^{2}\qlz^{4}.
\end{split}
\end{equation}
The values of the coefficients extracted from the fit are given in 
Table~\ref{tab:coefsY}.

\begin{table*}[htp]
\caption{\label{tab:coefsY}
  The harmonic, anharmonic, and nonlinear coupling terms extracted from the polynomial fit
  of the calculated total-energy surface $V(\qhy, \qlx, \qlz)$ of KTaO$_3$. The units are
  $\textrm{eV}/(\text{\AA}\sqrt{\textrm{u}})^{i+j+k}$, where $i$, $j$ and $k$ are
  the exponents of the phonon coordinates.}
\begin{ruledtabular}
\begin{tabular}{CCC|CCC}
  \text{Coefficient}  &\text{Order} & \text{Value} & \text{Coefficient}  &\text{Order} & \text{Value}\\
\hline
\olx^{2} & \qlx^{2} & 0.013628 & e_{6}& \qlx^{2}\qlz^{6} & 1.96\times10^{-7}\\
\olz^{2} & \qlz^{2} & 0.013628 & h_{1}& \qhy^{2}\qlx^{2} & -5.4947\times10^{-3}\\
\ohy^{2} & \qhy^{2} & 2.612711 & h_{2}& \qhy^{4}\qlx^{2} & -4.27\times10^{-4}\\

a_{4}& \qlx^{4} & 7.99\times10^{-4} & h_{3}& \qhy^{2}\qlx^{4} & 1.068\times10^{-5}\\
a_{6}& \qlx^{6} & -8.6\times10^{-6} & h_{4}& \qhy^{6}\qlx^{2} & -5.59\times10^{-6}\\
a_{8}& \qlx^{8} & 1.93\times10^{-7} & h_{5}& \qhy^{4}\qlx^{4} & 9.9\times10^{-7}\\

b_{4}& \qlz^{4} & 7.99\times10^{-4} & h_{6}& \qhy^{2}\qlx^{6} & 2.5\times10^{-7}\\
b_{6}& \qlz^{6} & -8.6\times10^{-6} & i_{1}& \qhy^{2}\qlz^{2} & -5.4947\times10^{-3}\\
b_{8}& \qlz^{8} & 1.93\times10^{-7} & i_{2}& \qhy^{4}\qlz^{2} & -4.27\times10^{-4}\\

d_{4}& \qhy^{4} & 0.06897 & i_{3}& \qhy^{2}\qlz^{4} & 1.068\times10^{-5}\\
d_{6}& \qhy^{6} & 7.8\times10^{-4} & i_{4}& \qhy^{6}\qlz^{2} & -5.59\times10^{-6}\\
d_{8}& \qhy^{8} & -1.1\times10^{-5} & i_{5}& \qhy^{4}\qlz^{4} & 9.9\times10^{-7}\\

e_{1}& \qlx^{2}\qlz^{2} & 2.796\times10^{-4} & i_{6}& \qhy^{2}\qlz^{6} & 2.5\times10^{-7}\\
e_{2}& \qlx^{4}\qlz^{2} & -1.14\times10^{-5} & k_{1}& \qhy^{2}\qlx^{2}\qlz^{2} & 3.83\times10^{-5}\\
e_{3}& \qlx^{2}\qlz^{4} & -1.14\times10^{-5} & k_{2}& \qhy^{4}\qlx^{2}\qlz^{2} & 6.1\times10^{-6}\\
e_{4}& \qlx^{6}\qlz^{2} & 1.96\times10^{-7} & k_{3}& \qhy^{2}\qlx^{4}\qlz^{2} & -3.6\times10^{-7}\\
e_{5}& \qlx^{4}\qlz^{4} & 2.35\times10^{-7} & k_{4}& \qhy^{2}\qlx^{2}\qlz^{4} & -3.6\times10^{-7}\\

\end{tabular}
\end{ruledtabular}
\end{table*}

\twocolumngrid

\bibliography{article}

\providecommand{\noopsort}[1]{}\providecommand{\singleletter}[1]{#1}%
\begin{thebibliography}{43}%
\makeatletter
\providecommand \@ifxundefined [1]{%
 \@ifx{#1\undefined}
}%
\providecommand \@ifnum [1]{%
 \ifnum #1\expandafter \@firstoftwo
 \else \expandafter \@secondoftwo
 \fi
}%
\providecommand \@ifx [1]{%
 \ifx #1\expandafter \@firstoftwo
 \else \expandafter \@secondoftwo
 \fi
}%
\providecommand \natexlab [1]{#1}%
\providecommand \enquote  [1]{``#1''}%
\providecommand \bibnamefont  [1]{#1}%
\providecommand \bibfnamefont [1]{#1}%
\providecommand \citenamefont [1]{#1}%
\providecommand \href@noop [0]{\@secondoftwo}%
\providecommand \href [0]{\begingroup \@sanitize@url \@href}%
\providecommand \@href[1]{\@@startlink{#1}\@@href}%
\providecommand \@@href[1]{\endgroup#1\@@endlink}%
\providecommand \@sanitize@url [0]{\catcode `\\12\catcode `\$12\catcode
  `\&12\catcode `\#12\catcode `\^12\catcode `\_12\catcode `\%12\relax}%
\providecommand \@@startlink[1]{}%
\providecommand \@@endlink[0]{}%
\providecommand \url  [0]{\begingroup\@sanitize@url \@url }%
\providecommand \@url [1]{\endgroup\@href {#1}{\urlprefix }}%
\providecommand \urlprefix  [0]{URL }%
\providecommand \Eprint [0]{\href }%
\providecommand \doibase [0]{https://doi.org/}%
\providecommand \selectlanguage [0]{\@gobble}%
\providecommand \bibinfo  [0]{\@secondoftwo}%
\providecommand \bibfield  [0]{\@secondoftwo}%
\providecommand \translation [1]{[#1]}%
\providecommand \BibitemOpen [0]{}%
\providecommand \bibitemStop [0]{}%
\providecommand \bibitemNoStop [0]{.\EOS\space}%
\providecommand \EOS [0]{\spacefactor3000\relax}%
\providecommand \BibitemShut  [1]{\csname bibitem#1\endcsname}%
\let\auto@bib@innerbib\@empty
\bibitem [{\citenamefont {Ovshinsky}(1968)}]{Ovshinsky1968}%
  \BibitemOpen
  \bibfield  {author} {\bibinfo {author} {\bibfnamefont {S.~R.}\ \bibnamefont
  {Ovshinsky}},\ }\href@noop {} {\bibfield  {journal} {\bibinfo  {journal}
  {Physical review letters}\ }\textbf {\bibinfo {volume} {21}},\ \bibinfo
  {pages} {1450} (\bibinfo {year} {1968})}\BibitemShut {NoStop}%
\bibitem [{\citenamefont {Stojchevska}\ \emph {et~al.}(2014)\citenamefont
  {Stojchevska}, \citenamefont {Vaskivskyi}, \citenamefont {Mertelj},
  \citenamefont {Kusar}, \citenamefont {Svetin}, \citenamefont {Brazovskii},\
  and\ \citenamefont {Mihailovic}}]{Stojchevska2014}%
  \BibitemOpen
  \bibfield  {author} {\bibinfo {author} {\bibfnamefont {L.}~\bibnamefont
  {Stojchevska}}, \bibinfo {author} {\bibfnamefont {I.}~\bibnamefont
  {Vaskivskyi}}, \bibinfo {author} {\bibfnamefont {T.}~\bibnamefont {Mertelj}},
  \bibinfo {author} {\bibfnamefont {P.}~\bibnamefont {Kusar}}, \bibinfo
  {author} {\bibfnamefont {D.}~\bibnamefont {Svetin}}, \bibinfo {author}
  {\bibfnamefont {S.}~\bibnamefont {Brazovskii}},\ and\ \bibinfo {author}
  {\bibfnamefont {D.}~\bibnamefont {Mihailovic}},\ }\href@noop {} {\bibfield
  {journal} {\bibinfo  {journal} {Science}\ }\textbf {\bibinfo {volume}
  {344}},\ \bibinfo {pages} {177} (\bibinfo {year} {2014})}\BibitemShut
  {NoStop}%
\bibitem [{\citenamefont {F{\"o}rst}\ \emph {et~al.}(2011)\citenamefont
  {F{\"o}rst}, \citenamefont {Manzoni}, \citenamefont {Kaiser}, \citenamefont
  {Tomioka}, \citenamefont {Tokura}, \citenamefont {Merlin},\ and\
  \citenamefont {Cavalleri}}]{Forst2011}%
  \BibitemOpen
  \bibfield  {author} {\bibinfo {author} {\bibfnamefont {M.}~\bibnamefont
  {F{\"o}rst}}, \bibinfo {author} {\bibfnamefont {C.}~\bibnamefont {Manzoni}},
  \bibinfo {author} {\bibfnamefont {S.}~\bibnamefont {Kaiser}}, \bibinfo
  {author} {\bibfnamefont {Y.}~\bibnamefont {Tomioka}}, \bibinfo {author}
  {\bibfnamefont {Y.}~\bibnamefont {Tokura}}, \bibinfo {author} {\bibfnamefont
  {R.}~\bibnamefont {Merlin}},\ and\ \bibinfo {author} {\bibfnamefont
  {A.}~\bibnamefont {Cavalleri}},\ }\href {https://doi.org/10.1038/nphys2055}
  {\bibfield  {journal} {\bibinfo  {journal} {Nature Physics}\ }\textbf
  {\bibinfo {volume} {7}},\ \bibinfo {pages} {854} (\bibinfo {year}
  {2011})}\BibitemShut {NoStop}%
\bibitem [{\citenamefont {Subedi}(2015)}]{Subedi2015}%
  \BibitemOpen
  \bibfield  {author} {\bibinfo {author} {\bibfnamefont {A.}~\bibnamefont
  {Subedi}},\ }\href {https://doi.org/10.1103/PhysRevB.92.214303} {\bibfield
  {journal} {\bibinfo  {journal} {Phys. Rev. B}\ }\textbf {\bibinfo {volume}
  {92}},\ \bibinfo {pages} {214303} (\bibinfo {year} {2015})}\BibitemShut
  {NoStop}%
\bibitem [{\citenamefont {Mankowsky}\ \emph {et~al.}(2017)\citenamefont
  {Mankowsky}, \citenamefont {von Hoegen}, \citenamefont {F\"orst},\ and\
  \citenamefont {Cavalleri}}]{Mankowsky2017}%
  \BibitemOpen
  \bibfield  {author} {\bibinfo {author} {\bibfnamefont {R.}~\bibnamefont
  {Mankowsky}}, \bibinfo {author} {\bibfnamefont {A.}~\bibnamefont {von
  Hoegen}}, \bibinfo {author} {\bibfnamefont {M.}~\bibnamefont {F\"orst}},\
  and\ \bibinfo {author} {\bibfnamefont {A.}~\bibnamefont {Cavalleri}},\ }\href
  {https://doi.org/10.1103/PhysRevLett.118.197601} {\bibfield  {journal}
  {\bibinfo  {journal} {Phys. Rev. Lett.}\ }\textbf {\bibinfo {volume} {118}},\
  \bibinfo {pages} {197601} (\bibinfo {year} {2017})}\BibitemShut {NoStop}%
\bibitem [{\citenamefont {Henstridge}\ \emph {et~al.}(2022)\citenamefont
  {Henstridge}, \citenamefont {F{\"o}rst}, \citenamefont {Rowe}, \citenamefont
  {Fechner},\ and\ \citenamefont {Cavalleri}}]{Henstridge2022}%
  \BibitemOpen
  \bibfield  {author} {\bibinfo {author} {\bibfnamefont {M.}~\bibnamefont
  {Henstridge}}, \bibinfo {author} {\bibfnamefont {M.}~\bibnamefont
  {F{\"o}rst}}, \bibinfo {author} {\bibfnamefont {E.}~\bibnamefont {Rowe}},
  \bibinfo {author} {\bibfnamefont {M.}~\bibnamefont {Fechner}},\ and\ \bibinfo
  {author} {\bibfnamefont {A.}~\bibnamefont {Cavalleri}},\ }\href@noop {}
  {\bibfield  {journal} {\bibinfo  {journal} {Nature Physics}\ }\textbf
  {\bibinfo {volume} {18}},\ \bibinfo {pages} {457} (\bibinfo {year}
  {2022})}\BibitemShut {NoStop}%
\bibitem [{\citenamefont {Mertelj}\ and\ \citenamefont
  {Kabanov}(2019)}]{Mertelj2019}%
  \BibitemOpen
  \bibfield  {author} {\bibinfo {author} {\bibfnamefont {T.}~\bibnamefont
  {Mertelj}}\ and\ \bibinfo {author} {\bibfnamefont {V.~V.}\ \bibnamefont
  {Kabanov}},\ }\href {https://doi.org/10.1103/PhysRevLett.123.129701}
  {\bibfield  {journal} {\bibinfo  {journal} {Phys. Rev. Lett.}\ }\textbf
  {\bibinfo {volume} {123}},\ \bibinfo {pages} {129701} (\bibinfo {year}
  {2019})}\BibitemShut {NoStop}%
\bibitem [{\citenamefont {Abalmasov}(2020)}]{Abalmasov2020}%
  \BibitemOpen
  \bibfield  {author} {\bibinfo {author} {\bibfnamefont {V.~A.}\ \bibnamefont
  {Abalmasov}},\ }\href {https://doi.org/10.1103/PhysRevB.101.014102}
  {\bibfield  {journal} {\bibinfo  {journal} {Phys. Rev. B}\ }\textbf {\bibinfo
  {volume} {101}},\ \bibinfo {pages} {014102} (\bibinfo {year}
  {2020})}\BibitemShut {NoStop}%
\bibitem [{\citenamefont {Chen}\ \emph {et~al.}(2022)\citenamefont {Chen},
  \citenamefont {Paillard}, \citenamefont {Zhao}, \citenamefont
  {{\'I}{\~{n}}iguez},\ and\ \citenamefont {Bellaiche}}]{Chen2022}%
  \BibitemOpen
  \bibfield  {author} {\bibinfo {author} {\bibfnamefont {P.}~\bibnamefont
  {Chen}}, \bibinfo {author} {\bibfnamefont {C.}~\bibnamefont {Paillard}},
  \bibinfo {author} {\bibfnamefont {H.~J.}\ \bibnamefont {Zhao}}, \bibinfo
  {author} {\bibfnamefont {J.}~\bibnamefont {{\'I}{\~{n}}iguez}},\ and\
  \bibinfo {author} {\bibfnamefont {L.}~\bibnamefont {Bellaiche}},\ }\href
  {https://doi.org/10.1038/s41467-022-30324-5} {\bibfield  {journal} {\bibinfo
  {journal} {Nature Communications}\ }\textbf {\bibinfo {volume} {13}},\
  \bibinfo {pages} {2566} (\bibinfo {year} {2022})}\BibitemShut {NoStop}%
\bibitem [{\citenamefont {Subedi}\ \emph {et~al.}(2014)\citenamefont {Subedi},
  \citenamefont {Cavalleri},\ and\ \citenamefont {Georges}}]{Subedi2014}%
  \BibitemOpen
  \bibfield  {author} {\bibinfo {author} {\bibfnamefont {A.}~\bibnamefont
  {Subedi}}, \bibinfo {author} {\bibfnamefont {A.}~\bibnamefont {Cavalleri}},\
  and\ \bibinfo {author} {\bibfnamefont {A.}~\bibnamefont {Georges}},\ }\href
  {https://doi.org/10.1103/PhysRevB.89.220301} {\bibfield  {journal} {\bibinfo
  {journal} {Phys. Rev. B}\ }\textbf {\bibinfo {volume} {89}},\ \bibinfo
  {pages} {220301} (\bibinfo {year} {2014})}\BibitemShut {NoStop}%
\bibitem [{\citenamefont {Subedi}(2017)}]{Subedi2017}%
  \BibitemOpen
  \bibfield  {author} {\bibinfo {author} {\bibfnamefont {A.}~\bibnamefont
  {Subedi}},\ }\href {https://doi.org/10.1103/PhysRevB.95.134113} {\bibfield
  {journal} {\bibinfo  {journal} {Phys. Rev. B}\ }\textbf {\bibinfo {volume}
  {95}},\ \bibinfo {pages} {134113} (\bibinfo {year} {2017})}\BibitemShut
  {NoStop}%
\bibitem [{\citenamefont {Radaelli}(2018)}]{Radaelli2018}%
  \BibitemOpen
  \bibfield  {author} {\bibinfo {author} {\bibfnamefont {P.~G.}\ \bibnamefont
  {Radaelli}},\ }\href@noop {} {\bibfield  {journal} {\bibinfo  {journal}
  {Physical Review B}\ }\textbf {\bibinfo {volume} {97}},\ \bibinfo {pages}
  {085145} (\bibinfo {year} {2018})}\BibitemShut {NoStop}%
\bibitem [{\citenamefont {Melnikov}\ \emph {et~al.}(2020)\citenamefont
  {Melnikov}, \citenamefont {Selivanov},\ and\ \citenamefont
  {Chekalin}}]{Melnikov2020}%
  \BibitemOpen
  \bibfield  {author} {\bibinfo {author} {\bibfnamefont {A.}~\bibnamefont
  {Melnikov}}, \bibinfo {author} {\bibfnamefont {Y.~G.}\ \bibnamefont
  {Selivanov}},\ and\ \bibinfo {author} {\bibfnamefont {S.}~\bibnamefont
  {Chekalin}},\ }\href@noop {} {\bibfield  {journal} {\bibinfo  {journal}
  {Physical Review B}\ }\textbf {\bibinfo {volume} {102}},\ \bibinfo {pages}
  {224301} (\bibinfo {year} {2020})}\BibitemShut {NoStop}%
\bibitem [{\citenamefont {Gu}\ and\ \citenamefont {Rondinelli}(2016)}]{Gu2016}%
  \BibitemOpen
  \bibfield  {author} {\bibinfo {author} {\bibfnamefont {M.}~\bibnamefont
  {Gu}}\ and\ \bibinfo {author} {\bibfnamefont {J.~M.}\ \bibnamefont
  {Rondinelli}},\ }\href@noop {} {\bibfield  {journal} {\bibinfo  {journal}
  {Scientific reports}\ }\textbf {\bibinfo {volume} {6}},\ \bibinfo {pages}
  {25121} (\bibinfo {year} {2016})}\BibitemShut {NoStop}%
\bibitem [{\citenamefont {Juraschek}\ \emph {et~al.}(2017)\citenamefont
  {Juraschek}, \citenamefont {Fechner}, \citenamefont {Balatsky},\ and\
  \citenamefont {Spaldin}}]{Juraschek2017b}%
  \BibitemOpen
  \bibfield  {author} {\bibinfo {author} {\bibfnamefont {D.~M.}\ \bibnamefont
  {Juraschek}}, \bibinfo {author} {\bibfnamefont {M.}~\bibnamefont {Fechner}},
  \bibinfo {author} {\bibfnamefont {A.~V.}\ \bibnamefont {Balatsky}},\ and\
  \bibinfo {author} {\bibfnamefont {N.~A.}\ \bibnamefont {Spaldin}},\
  }\href@noop {} {\bibfield  {journal} {\bibinfo  {journal} {Physical Review
  Materials}\ }\textbf {\bibinfo {volume} {1}},\ \bibinfo {pages} {014401}
  (\bibinfo {year} {2017})}\BibitemShut {NoStop}%
\bibitem [{\citenamefont {Fechner}\ \emph {et~al.}(2018)\citenamefont
  {Fechner}, \citenamefont {Sukhov}, \citenamefont {Chotorlishvili},
  \citenamefont {Kenel}, \citenamefont {Berakdar},\ and\ \citenamefont
  {Spaldin}}]{Fechner2018}%
  \BibitemOpen
  \bibfield  {author} {\bibinfo {author} {\bibfnamefont {M.}~\bibnamefont
  {Fechner}}, \bibinfo {author} {\bibfnamefont {A.}~\bibnamefont {Sukhov}},
  \bibinfo {author} {\bibfnamefont {L.}~\bibnamefont {Chotorlishvili}},
  \bibinfo {author} {\bibfnamefont {C.}~\bibnamefont {Kenel}}, \bibinfo
  {author} {\bibfnamefont {J.}~\bibnamefont {Berakdar}},\ and\ \bibinfo
  {author} {\bibfnamefont {N.}~\bibnamefont {Spaldin}},\ }\href@noop {}
  {\bibfield  {journal} {\bibinfo  {journal} {Physical review materials}\
  }\textbf {\bibinfo {volume} {2}},\ \bibinfo {pages} {064401} (\bibinfo {year}
  {2018})}\BibitemShut {NoStop}%
\bibitem [{\citenamefont {Khalsa}\ and\ \citenamefont
  {Benedek}(2018)}]{Khalsa2018}%
  \BibitemOpen
  \bibfield  {author} {\bibinfo {author} {\bibfnamefont {G.}~\bibnamefont
  {Khalsa}}\ and\ \bibinfo {author} {\bibfnamefont {N.~A.}\ \bibnamefont
  {Benedek}},\ }\href@noop {} {\bibfield  {journal} {\bibinfo  {journal} {npj
  Quantum Materials}\ }\textbf {\bibinfo {volume} {3}},\ \bibinfo {pages} {1}
  (\bibinfo {year} {2018})}\BibitemShut {NoStop}%
\bibitem [{\citenamefont {Gu}\ and\ \citenamefont {Rondinelli}(2018)}]{Gu2018}%
  \BibitemOpen
  \bibfield  {author} {\bibinfo {author} {\bibfnamefont {M.}~\bibnamefont
  {Gu}}\ and\ \bibinfo {author} {\bibfnamefont {J.~M.}\ \bibnamefont
  {Rondinelli}},\ }\href@noop {} {\bibfield  {journal} {\bibinfo  {journal}
  {Physical Review B}\ }\textbf {\bibinfo {volume} {98}},\ \bibinfo {pages}
  {024102} (\bibinfo {year} {2018})}\BibitemShut {NoStop}%
\bibitem [{\citenamefont {Rodriguez-Vega}\ \emph {et~al.}(2022)\citenamefont
  {Rodriguez-Vega}, \citenamefont {Lin}, \citenamefont {Leonardo},
  \citenamefont {Ernst}, \citenamefont {Vergniory},\ and\ \citenamefont
  {Fiete}}]{Rodriguez2022}%
  \BibitemOpen
  \bibfield  {author} {\bibinfo {author} {\bibfnamefont {M.}~\bibnamefont
  {Rodriguez-Vega}}, \bibinfo {author} {\bibfnamefont {Z.-X.}\ \bibnamefont
  {Lin}}, \bibinfo {author} {\bibfnamefont {A.}~\bibnamefont {Leonardo}},
  \bibinfo {author} {\bibfnamefont {A.}~\bibnamefont {Ernst}}, \bibinfo
  {author} {\bibfnamefont {M.~G.}\ \bibnamefont {Vergniory}},\ and\ \bibinfo
  {author} {\bibfnamefont {G.~A.}\ \bibnamefont {Fiete}},\ }\href@noop {}
  {\bibfield  {journal} {\bibinfo  {journal} {The Journal of Physical Chemistry
  Letters}\ }\textbf {\bibinfo {volume} {13}},\ \bibinfo {pages} {4152}
  (\bibinfo {year} {2022})}\BibitemShut {NoStop}%
\bibitem [{\citenamefont {Juraschek}\ \emph {et~al.}(2022)\citenamefont
  {Juraschek}, \citenamefont {Neuman},\ and\ \citenamefont
  {Narang}}]{Juraschek2022}%
  \BibitemOpen
  \bibfield  {author} {\bibinfo {author} {\bibfnamefont {D.~M.}\ \bibnamefont
  {Juraschek}}, \bibinfo {author} {\bibfnamefont {T.}~\bibnamefont {Neuman}},\
  and\ \bibinfo {author} {\bibfnamefont {P.}~\bibnamefont {Narang}},\
  }\href@noop {} {\bibfield  {journal} {\bibinfo  {journal} {Physical Review
  Research}\ }\textbf {\bibinfo {volume} {4}},\ \bibinfo {pages} {013129}
  (\bibinfo {year} {2022})}\BibitemShut {NoStop}%
\bibitem [{\citenamefont {Feng}\ \emph {et~al.}(2022)\citenamefont {Feng},
  \citenamefont {Han}, \citenamefont {Lan}, \citenamefont {Xu}, \citenamefont
  {Bi}, \citenamefont {Lin},\ and\ \citenamefont {Nan}}]{Feng2022}%
  \BibitemOpen
  \bibfield  {author} {\bibinfo {author} {\bibfnamefont {N.}~\bibnamefont
  {Feng}}, \bibinfo {author} {\bibfnamefont {J.}~\bibnamefont {Han}}, \bibinfo
  {author} {\bibfnamefont {C.}~\bibnamefont {Lan}}, \bibinfo {author}
  {\bibfnamefont {B.}~\bibnamefont {Xu}}, \bibinfo {author} {\bibfnamefont
  {K.}~\bibnamefont {Bi}}, \bibinfo {author} {\bibfnamefont {Y.}~\bibnamefont
  {Lin}},\ and\ \bibinfo {author} {\bibfnamefont {C.}~\bibnamefont {Nan}},\
  }\href@noop {} {\bibfield  {journal} {\bibinfo  {journal} {Physical Review
  B}\ }\textbf {\bibinfo {volume} {105}},\ \bibinfo {pages} {024304} (\bibinfo
  {year} {2022})}\BibitemShut {NoStop}%
\bibitem [{\citenamefont {Nova}\ \emph {et~al.}(2017)\citenamefont {Nova},
  \citenamefont {Cartella}, \citenamefont {Cantaluppi}, \citenamefont
  {F{\"o}rst}, \citenamefont {Bossini}, \citenamefont {Mikhaylovskiy},
  \citenamefont {Kimel}, \citenamefont {Merlin},\ and\ \citenamefont
  {Cavalleri}}]{Nova2017}%
  \BibitemOpen
  \bibfield  {author} {\bibinfo {author} {\bibfnamefont {T.~F.}\ \bibnamefont
  {Nova}}, \bibinfo {author} {\bibfnamefont {A.}~\bibnamefont {Cartella}},
  \bibinfo {author} {\bibfnamefont {A.}~\bibnamefont {Cantaluppi}}, \bibinfo
  {author} {\bibfnamefont {M.}~\bibnamefont {F{\"o}rst}}, \bibinfo {author}
  {\bibfnamefont {D.}~\bibnamefont {Bossini}}, \bibinfo {author} {\bibfnamefont
  {R.~V.}\ \bibnamefont {Mikhaylovskiy}}, \bibinfo {author} {\bibfnamefont
  {A.~V.}\ \bibnamefont {Kimel}}, \bibinfo {author} {\bibfnamefont
  {R.}~\bibnamefont {Merlin}},\ and\ \bibinfo {author} {\bibfnamefont
  {A.}~\bibnamefont {Cavalleri}},\ }\href@noop {} {\bibfield  {journal}
  {\bibinfo  {journal} {Nature Physics}\ }\textbf {\bibinfo {volume} {13}},\
  \bibinfo {pages} {132} (\bibinfo {year} {2017})}\BibitemShut {NoStop}%
\bibitem [{\citenamefont {Disa}\ \emph {et~al.}(2020)\citenamefont {Disa},
  \citenamefont {Fechner}, \citenamefont {Nova}, \citenamefont {Liu},
  \citenamefont {F{\"o}rst}, \citenamefont {Prabhakaran}, \citenamefont
  {Radaelli},\ and\ \citenamefont {Cavalleri}}]{Disa2020}%
  \BibitemOpen
  \bibfield  {author} {\bibinfo {author} {\bibfnamefont {A.~S.}\ \bibnamefont
  {Disa}}, \bibinfo {author} {\bibfnamefont {M.}~\bibnamefont {Fechner}},
  \bibinfo {author} {\bibfnamefont {T.~F.}\ \bibnamefont {Nova}}, \bibinfo
  {author} {\bibfnamefont {B.}~\bibnamefont {Liu}}, \bibinfo {author}
  {\bibfnamefont {M.}~\bibnamefont {F{\"o}rst}}, \bibinfo {author}
  {\bibfnamefont {D.}~\bibnamefont {Prabhakaran}}, \bibinfo {author}
  {\bibfnamefont {P.~G.}\ \bibnamefont {Radaelli}},\ and\ \bibinfo {author}
  {\bibfnamefont {A.}~\bibnamefont {Cavalleri}},\ }\href@noop {} {\bibfield
  {journal} {\bibinfo  {journal} {Nature Physics}\ }\textbf {\bibinfo {volume}
  {16}},\ \bibinfo {pages} {937} (\bibinfo {year} {2020})}\BibitemShut
  {NoStop}%
\bibitem [{\citenamefont {Stupakiewicz}\ \emph {et~al.}(2021)\citenamefont
  {Stupakiewicz}, \citenamefont {Davies}, \citenamefont {Szerenos},
  \citenamefont {Afanasiev}, \citenamefont {Rabinovich}, \citenamefont {Boris},
  \citenamefont {Caviglia}, \citenamefont {Kimel},\ and\ \citenamefont
  {Kirilyuk}}]{Stupakiewicz2021}%
  \BibitemOpen
  \bibfield  {author} {\bibinfo {author} {\bibfnamefont {A.}~\bibnamefont
  {Stupakiewicz}}, \bibinfo {author} {\bibfnamefont {C.}~\bibnamefont
  {Davies}}, \bibinfo {author} {\bibfnamefont {K.}~\bibnamefont {Szerenos}},
  \bibinfo {author} {\bibfnamefont {D.}~\bibnamefont {Afanasiev}}, \bibinfo
  {author} {\bibfnamefont {K.}~\bibnamefont {Rabinovich}}, \bibinfo {author}
  {\bibfnamefont {A.}~\bibnamefont {Boris}}, \bibinfo {author} {\bibfnamefont
  {A.}~\bibnamefont {Caviglia}}, \bibinfo {author} {\bibfnamefont
  {A.}~\bibnamefont {Kimel}},\ and\ \bibinfo {author} {\bibfnamefont
  {A.}~\bibnamefont {Kirilyuk}},\ }\href@noop {} {\bibfield  {journal}
  {\bibinfo  {journal} {Nature Physics}\ }\textbf {\bibinfo {volume} {17}},\
  \bibinfo {pages} {489} (\bibinfo {year} {2021})}\BibitemShut {NoStop}%
\bibitem [{\citenamefont {Afanasiev}\ \emph {et~al.}(2021)\citenamefont
  {Afanasiev}, \citenamefont {Hortensius}, \citenamefont {Ivanov},
  \citenamefont {Sasani}, \citenamefont {Bousquet}, \citenamefont {Blanter},
  \citenamefont {Mikhaylovskiy}, \citenamefont {Kimel},\ and\ \citenamefont
  {Caviglia}}]{Afanasiev2021}%
  \BibitemOpen
  \bibfield  {author} {\bibinfo {author} {\bibfnamefont {D.}~\bibnamefont
  {Afanasiev}}, \bibinfo {author} {\bibfnamefont {J.}~\bibnamefont
  {Hortensius}}, \bibinfo {author} {\bibfnamefont {B.}~\bibnamefont {Ivanov}},
  \bibinfo {author} {\bibfnamefont {A.}~\bibnamefont {Sasani}}, \bibinfo
  {author} {\bibfnamefont {E.}~\bibnamefont {Bousquet}}, \bibinfo {author}
  {\bibfnamefont {Y.}~\bibnamefont {Blanter}}, \bibinfo {author} {\bibfnamefont
  {R.}~\bibnamefont {Mikhaylovskiy}}, \bibinfo {author} {\bibfnamefont
  {A.}~\bibnamefont {Kimel}},\ and\ \bibinfo {author} {\bibfnamefont
  {A.}~\bibnamefont {Caviglia}},\ }\href@noop {} {\bibfield  {journal}
  {\bibinfo  {journal} {Nature materials}\ }\textbf {\bibinfo {volume} {20}},\
  \bibinfo {pages} {607} (\bibinfo {year} {2021})}\BibitemShut {NoStop}%
\bibitem [{\citenamefont {Disa}\ \emph {et~al.}(2023)\citenamefont {Disa},
  \citenamefont {Curtis}, \citenamefont {Fechner}, \citenamefont {Liu},
  \citenamefont {von Hoegen}, \citenamefont {F{\"o}rst}, \citenamefont {Nova},
  \citenamefont {Narang}, \citenamefont {Maljuk}, \citenamefont {Boris},
  \citenamefont {Keimer},\ and\ \citenamefont {Cavalleri}}]{Disa2023}%
  \BibitemOpen
  \bibfield  {author} {\bibinfo {author} {\bibfnamefont {A.~S.}\ \bibnamefont
  {Disa}}, \bibinfo {author} {\bibfnamefont {J.}~\bibnamefont {Curtis}},
  \bibinfo {author} {\bibfnamefont {M.}~\bibnamefont {Fechner}}, \bibinfo
  {author} {\bibfnamefont {A.}~\bibnamefont {Liu}}, \bibinfo {author}
  {\bibfnamefont {A.}~\bibnamefont {von Hoegen}}, \bibinfo {author}
  {\bibfnamefont {M.}~\bibnamefont {F{\"o}rst}}, \bibinfo {author}
  {\bibfnamefont {T.~F.}\ \bibnamefont {Nova}}, \bibinfo {author}
  {\bibfnamefont {P.}~\bibnamefont {Narang}}, \bibinfo {author} {\bibfnamefont
  {A.}~\bibnamefont {Maljuk}}, \bibinfo {author} {\bibfnamefont {A.~V.}\
  \bibnamefont {Boris}}, \bibinfo {author} {\bibfnamefont {B.}~\bibnamefont
  {Keimer}},\ and\ \bibinfo {author} {\bibfnamefont {A.}~\bibnamefont
  {Cavalleri}},\ }\href {https://doi.org/10.1038/s41586-023-05853-8} {\bibfield
   {journal} {\bibinfo  {journal} {Nature}\ }\textbf {\bibinfo {volume}
  {617}},\ \bibinfo {pages} {73} (\bibinfo {year} {2023})}\BibitemShut
  {NoStop}%
\bibitem [{\citenamefont {G\'omez~Pueyo}\ and\ \citenamefont
  {Subedi}(2022)}]{Adrian2022}%
  \BibitemOpen
  \bibfield  {author} {\bibinfo {author} {\bibfnamefont {A.}~\bibnamefont
  {G\'omez~Pueyo}}\ and\ \bibinfo {author} {\bibfnamefont {A.}~\bibnamefont
  {Subedi}},\ }\href {https://doi.org/10.1103/PhysRevB.106.214305} {\bibfield
  {journal} {\bibinfo  {journal} {Phys. Rev. B}\ }\textbf {\bibinfo {volume}
  {106}},\ \bibinfo {pages} {214305} (\bibinfo {year} {2022})}\BibitemShut
  {NoStop}%
\bibitem [{\citenamefont {Subedi}(2021)}]{Subedi2021}%
  \BibitemOpen
  \bibfield  {author} {\bibinfo {author} {\bibfnamefont {A.}~\bibnamefont
  {Subedi}},\ }\href {https://doi.org/10.5802/crphys.44} {\bibfield  {journal}
  {\bibinfo  {journal} {Comptes Rendus. Physique}\ }\textbf {\bibinfo {volume}
  {22}},\ \bibinfo {pages} {161} (\bibinfo {year} {2021})}\BibitemShut
  {NoStop}%
\bibitem [{\citenamefont {Cartella}\ \emph {et~al.}(2018)\citenamefont
  {Cartella}, \citenamefont {Nova}, \citenamefont {Fechner}, \citenamefont
  {Merlin},\ and\ \citenamefont {Cavalleri}}]{Cartella2018}%
  \BibitemOpen
  \bibfield  {author} {\bibinfo {author} {\bibfnamefont {A.}~\bibnamefont
  {Cartella}}, \bibinfo {author} {\bibfnamefont {T.~F.}\ \bibnamefont {Nova}},
  \bibinfo {author} {\bibfnamefont {M.}~\bibnamefont {Fechner}}, \bibinfo
  {author} {\bibfnamefont {R.}~\bibnamefont {Merlin}},\ and\ \bibinfo {author}
  {\bibfnamefont {A.}~\bibnamefont {Cavalleri}},\ }\href
  {https://doi.org/10.1073/pnas.1809725115} {\bibfield  {journal} {\bibinfo
  {journal} {Proceedings of the National Academy of Sciences}\ }\textbf
  {\bibinfo {volume} {115}},\ \bibinfo {pages} {12148} (\bibinfo {year}
  {2018})},\ \Eprint
  {https://arxiv.org/abs/https://www.pnas.org/content/115/48/12148.full.pdf}
  {https://www.pnas.org/content/115/48/12148.full.pdf} \BibitemShut {NoStop}%
\bibitem [{\citenamefont {Giannozzi}\ \emph {et~al.}(2020)\citenamefont
  {Giannozzi}, \citenamefont {Baseggio}, \citenamefont {Bonfà}, \citenamefont
  {Brunato}, \citenamefont {Car}, \citenamefont {Carnimeo}, \citenamefont
  {Cavazzoni}, \citenamefont {de~Gironcoli}, \citenamefont {Delugas},
  \citenamefont {Ferrari~Ruffino}, \citenamefont {Ferretti}, \citenamefont
  {Marzari}, \citenamefont {Timrov}, \citenamefont {Urru},\ and\ \citenamefont
  {Baroni}}]{QE}%
  \BibitemOpen
  \bibfield  {author} {\bibinfo {author} {\bibfnamefont {P.}~\bibnamefont
  {Giannozzi}}, \bibinfo {author} {\bibfnamefont {O.}~\bibnamefont {Baseggio}},
  \bibinfo {author} {\bibfnamefont {P.}~\bibnamefont {Bonfà}}, \bibinfo
  {author} {\bibfnamefont {D.}~\bibnamefont {Brunato}}, \bibinfo {author}
  {\bibfnamefont {R.}~\bibnamefont {Car}}, \bibinfo {author} {\bibfnamefont
  {I.}~\bibnamefont {Carnimeo}}, \bibinfo {author} {\bibfnamefont
  {C.}~\bibnamefont {Cavazzoni}}, \bibinfo {author} {\bibfnamefont
  {S.}~\bibnamefont {de~Gironcoli}}, \bibinfo {author} {\bibfnamefont
  {P.}~\bibnamefont {Delugas}}, \bibinfo {author} {\bibfnamefont
  {F.}~\bibnamefont {Ferrari~Ruffino}}, \bibinfo {author} {\bibfnamefont
  {A.}~\bibnamefont {Ferretti}}, \bibinfo {author} {\bibfnamefont
  {N.}~\bibnamefont {Marzari}}, \bibinfo {author} {\bibfnamefont
  {I.}~\bibnamefont {Timrov}}, \bibinfo {author} {\bibfnamefont
  {A.}~\bibnamefont {Urru}},\ and\ \bibinfo {author} {\bibfnamefont
  {S.}~\bibnamefont {Baroni}},\ }\href {https://doi.org/10.1063/5.0005082}
  {\bibfield  {journal} {\bibinfo  {journal} {The Journal of Chemical Physics}\
  }\textbf {\bibinfo {volume} {152}},\ \bibinfo {pages} {154105} (\bibinfo
  {year} {2020})},\ \Eprint
  {https://arxiv.org/abs/https://doi.org/10.1063/5.0005082}
  {https://doi.org/10.1063/5.0005082} \BibitemShut {NoStop}%
\bibitem [{\citenamefont {Perdew}\ \emph {et~al.}(2008)\citenamefont {Perdew},
  \citenamefont {Ruzsinszky}, \citenamefont {Csonka}, \citenamefont {Vydrov},
  \citenamefont {Scuseria}, \citenamefont {Constantin}, \citenamefont {Zhou},\
  and\ \citenamefont {Burke}}]{PBEsol}%
  \BibitemOpen
  \bibfield  {author} {\bibinfo {author} {\bibfnamefont {J.~P.}\ \bibnamefont
  {Perdew}}, \bibinfo {author} {\bibfnamefont {A.}~\bibnamefont {Ruzsinszky}},
  \bibinfo {author} {\bibfnamefont {G.~I.}\ \bibnamefont {Csonka}}, \bibinfo
  {author} {\bibfnamefont {O.~A.}\ \bibnamefont {Vydrov}}, \bibinfo {author}
  {\bibfnamefont {G.~E.}\ \bibnamefont {Scuseria}}, \bibinfo {author}
  {\bibfnamefont {L.~A.}\ \bibnamefont {Constantin}}, \bibinfo {author}
  {\bibfnamefont {X.}~\bibnamefont {Zhou}},\ and\ \bibinfo {author}
  {\bibfnamefont {K.}~\bibnamefont {Burke}},\ }\href
  {https://doi.org/10.1103/PhysRevLett.100.136406} {\bibfield  {journal}
  {\bibinfo  {journal} {Phys. Rev. Lett.}\ }\textbf {\bibinfo {volume} {100}},\
  \bibinfo {pages} {136406} (\bibinfo {year} {2008})}\BibitemShut {NoStop}%
\bibitem [{\citenamefont {Garrity}\ \emph {et~al.}(2014)\citenamefont
  {Garrity}, \citenamefont {Bennett}, \citenamefont {Rabe},\ and\ \citenamefont
  {Vanderbilt}}]{GBRV}%
  \BibitemOpen
  \bibfield  {author} {\bibinfo {author} {\bibfnamefont {K.~F.}\ \bibnamefont
  {Garrity}}, \bibinfo {author} {\bibfnamefont {J.~W.}\ \bibnamefont
  {Bennett}}, \bibinfo {author} {\bibfnamefont {K.~M.}\ \bibnamefont {Rabe}},\
  and\ \bibinfo {author} {\bibfnamefont {D.}~\bibnamefont {Vanderbilt}},\
  }\href {https://doi.org/https://doi.org/10.1016/j.commatsci.2013.08.053}
  {\bibfield  {journal} {\bibinfo  {journal} {Computational Materials Science}\
  }\textbf {\bibinfo {volume} {81}},\ \bibinfo {pages} {446} (\bibinfo {year}
  {2014})}\BibitemShut {NoStop}%
\bibitem [{\citenamefont {Savrasov}\ \emph {et~al.}(1994)\citenamefont
  {Savrasov}, \citenamefont {Savrasov},\ and\ \citenamefont
  {Andersen}}]{Savrasov1994}%
  \BibitemOpen
  \bibfield  {author} {\bibinfo {author} {\bibfnamefont {S.~Y.}\ \bibnamefont
  {Savrasov}}, \bibinfo {author} {\bibfnamefont {D.~Y.}\ \bibnamefont
  {Savrasov}},\ and\ \bibinfo {author} {\bibfnamefont {O.~K.}\ \bibnamefont
  {Andersen}},\ }\href {https://doi.org/10.1103/PhysRevLett.72.372} {\bibfield
  {journal} {\bibinfo  {journal} {Phys. Rev. Lett.}\ }\textbf {\bibinfo
  {volume} {72}},\ \bibinfo {pages} {372} (\bibinfo {year} {1994})}\BibitemShut
  {NoStop}%
\bibitem [{\citenamefont {Sokolowski-Tinten}\ \emph {et~al.}(2003)\citenamefont
  {Sokolowski-Tinten}, \citenamefont {Blome}, \citenamefont {Blums},
  \citenamefont {Cavalleri}, \citenamefont {Dietrich}, \citenamefont
  {Tarasevitch}, \citenamefont {Uschmann}, \citenamefont {F{\"o}rster},
  \citenamefont {Kammler}, \citenamefont {Horn-von Hoegen},\ and\ \citenamefont
  {von~der Linde}}]{Sokolowski2003}%
  \BibitemOpen
  \bibfield  {author} {\bibinfo {author} {\bibfnamefont {K.}~\bibnamefont
  {Sokolowski-Tinten}}, \bibinfo {author} {\bibfnamefont {C.}~\bibnamefont
  {Blome}}, \bibinfo {author} {\bibfnamefont {J.}~\bibnamefont {Blums}},
  \bibinfo {author} {\bibfnamefont {A.}~\bibnamefont {Cavalleri}}, \bibinfo
  {author} {\bibfnamefont {C.}~\bibnamefont {Dietrich}}, \bibinfo {author}
  {\bibfnamefont {A.}~\bibnamefont {Tarasevitch}}, \bibinfo {author}
  {\bibfnamefont {I.}~\bibnamefont {Uschmann}}, \bibinfo {author}
  {\bibfnamefont {E.}~\bibnamefont {F{\"o}rster}}, \bibinfo {author}
  {\bibfnamefont {M.}~\bibnamefont {Kammler}}, \bibinfo {author} {\bibfnamefont
  {M.}~\bibnamefont {Horn-von Hoegen}},\ and\ \bibinfo {author} {\bibfnamefont
  {D.}~\bibnamefont {von~der Linde}},\ }\href
  {https://doi.org/10.1038/nature01490} {\bibfield  {journal} {\bibinfo
  {journal} {Nature}\ }\textbf {\bibinfo {volume} {422}},\ \bibinfo {pages}
  {287} (\bibinfo {year} {2003})}\BibitemShut {NoStop}%
\bibitem [{\citenamefont {Bates}\ \emph {et~al.}(2022)\citenamefont {Bates},
  \citenamefont {Kornblith}, \citenamefont {Noack}, \citenamefont
  {Bouchet-Valat}, \citenamefont {Borregaard}, \citenamefont {Arslan},
  \citenamefont {White}, \citenamefont {Kleinschmidt}, \citenamefont {Lynch},
  \citenamefont {Dunning}, \citenamefont {Mogensen}, \citenamefont {Lendle},
  \citenamefont {Aluthge}, \citenamefont {pdeffebach}, \citenamefont {José
  Bayoán Santiago~Calderón}, \citenamefont {Born}, \citenamefont {Setzler},
  \citenamefont {DuBois}, \citenamefont {Quinn}, \citenamefont {Slámečka},
  \citenamefont {Bastide}, \citenamefont {Alday}, \citenamefont
  {Anthony~Blaom}, \citenamefont {König}, \citenamefont {Kamiński},
  \citenamefont {Caine}, \citenamefont {Lin},\ and\ \citenamefont
  {Karrasch}}]{GLM}%
  \BibitemOpen
  \bibfield  {author} {\bibinfo {author} {\bibfnamefont {D.}~\bibnamefont
  {Bates}}, \bibinfo {author} {\bibfnamefont {S.}~\bibnamefont {Kornblith}},
  \bibinfo {author} {\bibfnamefont {A.}~\bibnamefont {Noack}}, \bibinfo
  {author} {\bibfnamefont {M.}~\bibnamefont {Bouchet-Valat}}, \bibinfo {author}
  {\bibfnamefont {M.~K.}\ \bibnamefont {Borregaard}}, \bibinfo {author}
  {\bibfnamefont {A.}~\bibnamefont {Arslan}}, \bibinfo {author} {\bibfnamefont
  {J.~M.}\ \bibnamefont {White}}, \bibinfo {author} {\bibfnamefont
  {D.}~\bibnamefont {Kleinschmidt}}, \bibinfo {author} {\bibfnamefont
  {G.}~\bibnamefont {Lynch}}, \bibinfo {author} {\bibfnamefont
  {I.}~\bibnamefont {Dunning}}, \bibinfo {author} {\bibfnamefont {P.~K.}\
  \bibnamefont {Mogensen}}, \bibinfo {author} {\bibfnamefont {S.}~\bibnamefont
  {Lendle}}, \bibinfo {author} {\bibfnamefont {D.}~\bibnamefont {Aluthge}},
  \bibinfo {author} {\bibnamefont {pdeffebach}}, \bibinfo {author}
  {\bibfnamefont {P.}~\bibnamefont {José Bayoán Santiago~Calderón}},
  \bibinfo {author} {\bibfnamefont {B.}~\bibnamefont {Born}}, \bibinfo {author}
  {\bibfnamefont {B.}~\bibnamefont {Setzler}}, \bibinfo {author} {\bibfnamefont
  {C.}~\bibnamefont {DuBois}}, \bibinfo {author} {\bibfnamefont
  {J.}~\bibnamefont {Quinn}}, \bibinfo {author} {\bibfnamefont
  {O.}~\bibnamefont {Slámečka}}, \bibinfo {author} {\bibfnamefont
  {P.}~\bibnamefont {Bastide}}, \bibinfo {author} {\bibfnamefont
  {P.}~\bibnamefont {Alday}}, \bibinfo {author} {\bibfnamefont
  {P.}~\bibnamefont {Anthony~Blaom}}, \bibinfo {author} {\bibfnamefont
  {B.}~\bibnamefont {König}}, \bibinfo {author} {\bibfnamefont
  {B.}~\bibnamefont {Kamiński}}, \bibinfo {author} {\bibfnamefont
  {C.}~\bibnamefont {Caine}}, \bibinfo {author} {\bibfnamefont
  {D.}~\bibnamefont {Lin}},\ and\ \bibinfo {author} {\bibfnamefont
  {D.}~\bibnamefont {Karrasch}},\ }\href
  {https://doi.org/10.5281/zenodo.5823359} {\bibinfo {title}
  {Juliastats/glm.jl: v1.6.0}} (\bibinfo {year} {2022})\BibitemShut {NoStop}%
\bibitem [{\citenamefont {Souza}\ \emph {et~al.}(2002)\citenamefont {Souza},
  \citenamefont {\'I\~niguez},\ and\ \citenamefont {Vanderbilt}}]{Souza2002}%
  \BibitemOpen
  \bibfield  {author} {\bibinfo {author} {\bibfnamefont {I.}~\bibnamefont
  {Souza}}, \bibinfo {author} {\bibfnamefont {J.}~\bibnamefont {\'I\~niguez}},\
  and\ \bibinfo {author} {\bibfnamefont {D.}~\bibnamefont {Vanderbilt}},\
  }\href {https://doi.org/10.1103/PhysRevLett.89.117602} {\bibfield  {journal}
  {\bibinfo  {journal} {Phys. Rev. Lett.}\ }\textbf {\bibinfo {volume} {89}},\
  \bibinfo {pages} {117602} (\bibinfo {year} {2002})}\BibitemShut {NoStop}%
\bibitem [{\citenamefont {Rackauckas}\ and\ \citenamefont
  {Nie}(2017)}]{DifferentialEquations}%
  \BibitemOpen
  \bibfield  {author} {\bibinfo {author} {\bibfnamefont {C.}~\bibnamefont
  {Rackauckas}}\ and\ \bibinfo {author} {\bibfnamefont {Q.}~\bibnamefont
  {Nie}},\ }\href {https://doi.org/10.5334/jors.151} {\bibfield  {journal}
  {\bibinfo  {journal} {The Journal of Open Research Software}\ }\textbf
  {\bibinfo {volume} {5}} (\bibinfo {year} {2017})},\ \bibinfo {note} {exported
  from https://app.dimensions.ai on 2019/05/05}\BibitemShut {NoStop}%
\bibitem [{\citenamefont {Caprini}\ \emph {et~al.}(2023)\citenamefont
  {Caprini}, \citenamefont {Löwen},\ and\ \citenamefont
  {Geilhufe}}]{Caprini2023}%
  \BibitemOpen
  \bibfield  {author} {\bibinfo {author} {\bibfnamefont {L.}~\bibnamefont
  {Caprini}}, \bibinfo {author} {\bibfnamefont {H.}~\bibnamefont {Löwen}},\
  and\ \bibinfo {author} {\bibfnamefont {R.~M.}\ \bibnamefont {Geilhufe}},\
  }\href@noop {} {\bibinfo {title} {Ultrafast entropy production in pump-probe
  experiments}} (\bibinfo {year} {2023}),\ \Eprint
  {https://arxiv.org/abs/2302.02716} {arXiv:2302.02716 [cond-mat.mtrl-sci]}
  \BibitemShut {NoStop}%
\bibitem [{\citenamefont {Juraschek}\ \emph {et~al.}(2020)\citenamefont
  {Juraschek}, \citenamefont {Meier},\ and\ \citenamefont
  {Narang}}]{Juraschek2020}%
  \BibitemOpen
  \bibfield  {author} {\bibinfo {author} {\bibfnamefont {D.~M.}\ \bibnamefont
  {Juraschek}}, \bibinfo {author} {\bibfnamefont {Q.~N.}\ \bibnamefont
  {Meier}},\ and\ \bibinfo {author} {\bibfnamefont {P.}~\bibnamefont
  {Narang}},\ }\href {https://doi.org/10.1103/PhysRevLett.124.117401}
  {\bibfield  {journal} {\bibinfo  {journal} {Phys. Rev. Lett.}\ }\textbf
  {\bibinfo {volume} {124}},\ \bibinfo {pages} {117401} (\bibinfo {year}
  {2020})}\BibitemShut {NoStop}%
\bibitem [{\citenamefont {Perry}\ \emph {et~al.}(1989)\citenamefont {Perry},
  \citenamefont {Currat}, \citenamefont {Buhay}, \citenamefont {Migoni},
  \citenamefont {Stirling},\ and\ \citenamefont {Axe}}]{Perry1989}%
  \BibitemOpen
  \bibfield  {author} {\bibinfo {author} {\bibfnamefont {C.~H.}\ \bibnamefont
  {Perry}}, \bibinfo {author} {\bibfnamefont {R.}~\bibnamefont {Currat}},
  \bibinfo {author} {\bibfnamefont {H.}~\bibnamefont {Buhay}}, \bibinfo
  {author} {\bibfnamefont {R.~M.}\ \bibnamefont {Migoni}}, \bibinfo {author}
  {\bibfnamefont {W.~G.}\ \bibnamefont {Stirling}},\ and\ \bibinfo {author}
  {\bibfnamefont {J.~D.}\ \bibnamefont {Axe}},\ }\href
  {https://doi.org/10.1103/PhysRevB.39.8666} {\bibfield  {journal} {\bibinfo
  {journal} {Phys. Rev. B}\ }\textbf {\bibinfo {volume} {39}},\ \bibinfo
  {pages} {8666} (\bibinfo {year} {1989})}\BibitemShut {NoStop}%
\bibitem [{\citenamefont {Nilsen}\ and\ \citenamefont
  {Skinner}(1967)}]{Nilsen1967}%
  \BibitemOpen
  \bibfield  {author} {\bibinfo {author} {\bibfnamefont {W.~G.}\ \bibnamefont
  {Nilsen}}\ and\ \bibinfo {author} {\bibfnamefont {J.~G.}\ \bibnamefont
  {Skinner}},\ }\href {https://doi.org/10.1063/1.1712096} {\bibfield  {journal}
  {\bibinfo  {journal} {The Journal of Chemical Physics}\ }\textbf {\bibinfo
  {volume} {47}},\ \bibinfo {pages} {1413} (\bibinfo {year} {1967})},\ \Eprint
  {https://arxiv.org/abs/https://doi.org/10.1063/1.1712096}
  {https://doi.org/10.1063/1.1712096} \BibitemShut {NoStop}%
\bibitem [{\citenamefont {Bartels}\ \emph {et~al.}(2000)\citenamefont
  {Bartels}, \citenamefont {Dekorsy},\ and\ \citenamefont
  {Kurz}}]{Bartels2000}%
  \BibitemOpen
  \bibfield  {author} {\bibinfo {author} {\bibfnamefont {A.}~\bibnamefont
  {Bartels}}, \bibinfo {author} {\bibfnamefont {T.}~\bibnamefont {Dekorsy}},\
  and\ \bibinfo {author} {\bibfnamefont {H.}~\bibnamefont {Kurz}},\ }\href
  {https://doi.org/10.1103/PhysRevLett.84.2981} {\bibfield  {journal} {\bibinfo
   {journal} {Phys. Rev. Lett.}\ }\textbf {\bibinfo {volume} {84}},\ \bibinfo
  {pages} {2981} (\bibinfo {year} {2000})}\BibitemShut {NoStop}%
\bibitem [{\citenamefont {Itin}\ and\ \citenamefont
  {Katsnelson}(2018)}]{Itin2018}%
  \BibitemOpen
  \bibfield  {author} {\bibinfo {author} {\bibfnamefont {A.}~\bibnamefont
  {Itin}}\ and\ \bibinfo {author} {\bibfnamefont {M.}~\bibnamefont
  {Katsnelson}},\ }\href@noop {} {\bibfield  {journal} {\bibinfo  {journal}
  {Physical Review B}\ }\textbf {\bibinfo {volume} {97}},\ \bibinfo {pages}
  {184304} (\bibinfo {year} {2018})}\BibitemShut {NoStop}%
\end{thebibliography}%

\end{document}